\documentstyle[12pt,epic]{article}
\catcode`\@=11

\font\teneus=eusm10 scaled \magstep1
\font\seveneus=eusm7 scaled \magstep1
\font\fiveeus=eusm5 scaled \magstep1

\newfam\eusfam
\textfont\eusfam=\teneus  \scriptfont\eusfam=\seveneus
  \scriptscriptfont\eusfam=\fiveeus

\def\hexnumber@#1{\ifnum#1<10 \number#1\else
 \ifnum#1=10 A\else\ifnum#1=11 B\else\ifnum#1=12 C\else
 \ifnum#1=13 D\else\ifnum#1=14 E\else\ifnum#1=15 F\fi\fi\fi\fi\fi\fi\fi}

\def\Cl{\ifmmode\let\next\Cl@\else
 \def\next{\errmessage{Use \string\Cl\space only in math mode}}\fi\next}
\def\Cl@#1{{\Cl@@{#1}}}
\def\Cl@@#1{\fam\eusfam#1}

\catcode`\@=11

\font\teneuf=eufm10 scaled \magstep1
\font\seveneuf=eufm7 scaled \magstep1
\font\fiveeuf=eufm5 scaled \magstep1

\newfam\euffam
\textfont\euffam=\teneuf  \scriptfont\euffam=\seveneuf
  \scriptscriptfont\euffam=\fiveeuf

\def\hexnumber@#1{\ifnum#1<10 \number#1\else
 \ifnum#1=10 A\else\ifnum#1=11 B\else\ifnum#1=12 C\else
 \ifnum#1=13 D\else\ifnum#1=14 E\else\ifnum#1=15 F\fi\fi\fi\fi\fi\fi\fi}

\def\Got{\ifmmode\let\next\Got@\else
 \def\next{\errmessage{Use \string\Got\space only in math mode}}\fi\next}
\def\Got@#1{{\Got@@{#1}}}
\def\Got@@#1{\fam\euffam#1}

\catcode`\@=11

\font\tenmsx=msxm10 scaled \magstep1
\font\sevenmsx=msxm7 scaled \magstep1
\font\fivemsx=msxm5 scaled \magstep1
\font\tenmsy=msym10 scaled \magstep1
\font\sevenmsy=msym7 scaled \magstep1
\font\fivemsy=msym5 scaled \magstep1
\newfam\msxfam
\newfam\msyfam
\textfont\msxfam=\tenmsx  \scriptfont\msxfam=\sevenmsx
  \scriptscriptfont\msxfam=\fivemsx
\textfont\msyfam=\tenmsy  \scriptfont\msyfam=\sevenmsy
  \scriptscriptfont\msyfam=\fivemsy

\def\hexnumber@#1{\ifnum#1<10 \number#1\else
 \ifnum#1=10 A\else\ifnum#1=11 B\else\ifnum#1=12 C\else
 \ifnum#1=13 D\else\ifnum#1=14 E\else\ifnum#1=15 F\fi\fi\fi\fi\fi\fi\fi}

\def\Bbb{\ifmmode\let\next\Bbb@\else
 \def\next{\errmessage{Use \string\Bbb\space only in math mode}}\fi\next}
\def\Bbb@#1{{\Bbb@@{#1}}}
\def\Bbb@@#1{\fam\msyfam#1}

\renewcommand{\theequation}{\thesection.\arabic{equation}}
\renewcommand{\title}[1]{\large\bf \mbox{}\\ \mbox{}\\ \mbox{}\\ \mbox{}\\
     #1\bigskip\medskip\\}
\renewcommand{\author}[1]{\large #1\\ \smallskip}

\newcommand{\hs}[1]{\mbox{\hspace*{#1cm}}}
\newcommand{\vs}[1]{\mbox{\vspace*{#1cm}}}

\newcommand{\address}[1]{{\narrower\normalsize\it #1\\}\bigskip}

\newcommand{\Z}{\mbox{\sf Z\hspace*{-0.45em}Z}}
\long\def\ignore#1{}
\def\ade{$A$--$D$--$E$\space}

\def\WF#1#2#3#4#5#6#7#8#9{W_{(a,p)}^{(b,q)}\mbox{$\left(
   \matrix{#5&#8&#4\cr#9&&#7\cr #1&#6&#3\cr}\Biggm|\mbox{$#2$}\right)$}}
\def\smat#1{\mbox{\small $\pmatrix{#1}$}}

\def\one{\hbox{{1}\kern-.25em\hbox{l}}}

\def\Or{\;\;{\rm or}\;\;}
\def\and{\;\;{\rm and}\;\;}

\def\re{\mbox{${\cal R}e$ }}
\def\i{i}
\def\-{\!-\!}
\def\+{\!+\!}
\def\<{\langle}
\def\>{\rangle}
\def\({\biggl(}
\def\){\biggr)}
\def\h{\hspace*{0.5cm}}
\def\mh{\hspace*{-0.5cm}}
\def\half {\mbox{$\textstyle {1 \over 2}$}}
\def\mat {\pmatrix}
\def\smat#1{\mbox{\small $\mat{#1}$}}
\def\ba{\begin{array}}
\def\ea{\end{array}}
\def\be{\begin{eqnarray}}
\def\ee{\end{eqnarray}}

\def\ol{\overline}
\def\no{\nonumber}

\def\T{\mbox{\boldmath { $T$}}}

\def\t{\mbox{\boldmath {$t$}}}
\def\s{\mbox{\boldmath {$s$}}}
\def\ro{\mbox{\boldmath $\rho$}}

\def\tt{\tilde{\t}}
\def\1{\mbox{\boldmath I}}

\def\a{\mbox{$\Got a$}}
\def\U{\mbox{$\Got  U$}}

\def\e{\mbox{$\Got e$}}
\def\b{\mbox{$\Got b$}}
\def\la{\mbox{${\ell}a$}}
\def\Le{\mbox{${\ell}e$}}
\def\lA{\mbox{${\ell}A$}}

\def\[{\mbox{$\biggl[$}}
\def\]{\mbox{$\biggr]$}}
\def\lf{\left\lfloor}
\def\rf{\right\rfloor}

\def\W{\mbox{$\cal W$}}
\long\def\twocol#1&#2\\{\par
\begin{minipage}[t]{2.9in}
\hspace{1cm}#1
\end{minipage}
\begin{minipage}[t]{2.9in}
\hspace{1cm}#2
\end{minipage}\smallskip\par}

\def\face#1#2#3#4#5{
  \begin{picture}(40,30)(-5,13)
  \thicklines
  \put(15,5){\makebox(0,0)[b]{\line(1,0){20}}}
  \put(5,5){\makebox(0,0)[b]{\line(0,1){20}}}
  \put(15,25){\makebox(0,0)[b]{\line(-1,0){20}}}
  \put(25,25){\makebox(0,0)[b]{\line(0,-1){20}}}
  \put(5,2){\makebox(0,0)[rb]{\sc $#1$}} \put(26,2){\sc $#2$}
  \put(25,25){\sc $#3$}\put(5,25){\makebox(0,0)[rb]{\sc $#4$}}
  \put(13,13){\sc $#5$}
 \end{picture}}

\begin{document}

\begin{flushright} 
\today
\end{flushright}
\begin{center}
\title{SOLUTION OF FUNCTIONAL EQUATIONS OF\\ RESTRICTED
            $A_{n-1}^{(1)}$  FUSED LATTICE MODELS}
\author{Yu-kui Zhou\footnote{Email: \mbox{ykzhou@mundoe.maths.mu.oz.au}}
and  Paul
A. Pearce\footnote{Email: \mbox{pap@mundoe.maths.mu.oz.au}}}
\address{Mathematics Department, University of Melbourne,\\Parkville, Victoria
3052, Australia }

\begin{abstract}
Functional equations, in the form of fusion hierarchies, are
studied for the transfer matrices of the fused restricted $A_{n-1}^{(1)}$
lattice models of Jimbo, Miwa and Okado. Specifically, these equations
are solved analytically for the finite-size  scaling spectra,
central charges and some conformal weights. The results are obtained
in terms of Rogers dilogarithm and correspond
to coset conformal field theories based on the affine Lie algebra
$A_{n-1}^{(1)}$ with GKO pair $A^{(1)}_{n-1}\;
\oplus A^{(1)}_{n-1}\;\supset \; A^{(1)}_{n-1}$.

\end{abstract}
\end{center}

\bigskip

hep-th/9502067

\bigskip

\section{Introduction}
\setcounter{equation}{0}

\subsection{Conformal spectra and lattice models}
In the last decade it has become clear that the critical behavior of
two-dimensional statistical systems can be described by conformal
field  theories \cite{BPZ:84,FQS:84,ABF:84,Huse:84}. This is possible
because statistical systems at critical points possess conformal invariance
\cite{Cardy:86}. In particular, the continuum limit of the  critical $L$-state
restricted  solid-on-solid (RSOS) models of Andrews, Baxter and
Forrester (ABF)~\cite{ABF:84} provide realizations of the unitary  minimal
conformal
field theories with central charges
\be
c\;=\;1- {6\over L(L+1)}
\ee
and conformal weights given by the Kac formula \cite{Kac:79}
\be
\Delta_{t,s}\;=\;{[(L+1)t-Ls]^2-1\over 4L(L+1)}\;,\h 1\le t\le L-1\;,\h 1\le
s\le L\;,\h s\le t\;.
\ee
At criticality, the characters of the Virasoro algebra of the conformal field
theory appear naturally in the modular invariant partition functions. In
particular,
the conformal spectra, including the central charge and the conformal
weights,  can
be obtained from the finite-size corrections to the transfer matrix
eigenvalues~\cite{BaRe:89,KlPe:92}.  Surprisingly, the same characters
appear~\cite{ABF:84,DJMO:87} in the expressions for the off-critical local
state
probabilities of the  ABF models.

Further conformal field theories are obtained from the fused ABF
RSOS models at fusion level $p$.  The central charge and conformal weights
of these theories are given by
\be
c&=&{3p\over p+2}-{6p\over (L+1)(L+1-p)}\; ,   \\
\Delta_{t,s}&=&{[(L+1)t-(L+1-p)s]^2-p^2\over 4p(L+1)(L+1-p)} +
    {s_0(p-s_0)\over 2p(p+2)}
\hs{1.5}\ee where
\be 0\;\le\; s_0 \;\le\; p \and s_0\;=\; \pm\; (t-s) \;\mbox{ mod $2p$}\; .
\ee
This conformal spectra was conjectured~\cite{DJKMO:87,DJKMO:88} by the
evaluation
of the  local state probabilities and confirmed by the direct
calculation~\cite{BaRe:89,KlPe:92} of finite-size corrections for the fused
ABF RSOS
models. The fused lattice models are related to coset conformal field theories
obtained by the Goddard-Kent-Olive (GKO) construction~\cite{GKO:85}. The
relevant
GKO pair is
\be
           & A^{(1)}_1 &\!\! \oplus\h A^{(1)}_1\;\supset \; A^{(1)}_1 \\ {\rm
level}&L-p-1      &\phantom{1opo}  p \phantom{11111}  L-1  \no\label{coset-2p}
\ee
The minimal unitary series corresponds to level  $p=1$. For $p>1$  the
conformal
characters are identified as branching coefficients.

Another direction to extend the unitary minimal series is to higher rank
theories
with the GKO pair
\be
& A^{(1)}_{n-1} &\!\! \oplus\h A^{(1)}_{n-1}\;\supset \; A^{(1)}_{n-1} \\
{\rm level}&l-1      &\phantom{1opoo}  1 \phantom{1111111}  l   \no
\label{coset-n}
\ee
These theories reduce to the  unitary minimal series for $n=2$ and $l=L-1$.
The extended Virasoro algebra $\W^n_1$ can be
constructed~\cite{FaLu:88,BBSS:88}
starting from  this GKO pair of the affine algebra
$A^{(1)}_{n-1}$. It has been confirmed that the integrable
lattice models corresponding to these higher rank coset conformal field
theories
(\ref{coset-n}) are the  $A^{(1)}_{n-1}$ lattice models
of Jimbo, Miwa and Okado~\cite{JMO:88}. The central charge and conformal
weights are
\be
c\;=\;(n-1)\left[1- {n(n+1)\over (l+n)(l+n-1)}\right] \label{c-n}
\ee
\be
\Delta_{\t,\s}\;=\;{[(l+n)\t-(l+n-1)\s]^2-
  n(n^2-1)/12\over 2(l+n-1)(l+n)}\;, \label{delta-n}
\ee
where $\t={\bf p}+\ro$ and $\s={\bf q}+\ro$.  Here $\ro$ is the Weyl vector or
the sum of all fundamental weights of $A^{(1)}_{n-1}$ and
${\bf p}$ and ${\bf q}$ are local states given respectively by the dominant
integral
weights  $P_+(n,l-1)$ of level $l-1$ and $P_+(n,l)$ of level $l$. These
notations
are briefly explained in subsection~\ref{notations}.  The conformal spectra
(\ref{c-n}) and (\ref{delta-n}) has been obtained by the GKO construction of
the
stress-energy in conformal field theory
\cite{ChRa:89} and the study of local state probabilities of  the JMO
lattice models
\cite{JMO:88}.

Further generalizing (\ref{coset-2p}) and (\ref{coset-n}) by the fusion
procedure
leads to the GKO pair
\be & A^{(1)}_{n-1} &\!\! \oplus\h A^{(1)}_{n-1}\;\supset \; A^{(1)}_{n-1}
\no \\
{\rm level}&l-p-1      &\phantom{1opoo}  p \phantom{1111111}  l
  \label{coset-np}
\ee
with Virasoro  algebra $\W^n_p$. Using the GKO construction of the
stress-energy
tensor, the central charge \cite{ChRa:89,BaRe:90,KNS:93,KNS:93b} and
conformal weights \cite{ChRa:89} of the $\W^n_p$ models are
given by
\be c\;=\;{(n^2-1)p\over p+n}- {n(n^2-1)p\over (l+n)(l+n-p)} \label{c-np}
\ee
\be
\Delta_{\t,\s}\;=\;{[(l+n)\t-(l+n-p)\s]^2- p^2n(n^2-1)/12
  \over 2p(l+n-p)(l+n)} +\delta_{_{\mbox{\boldmath$\nu$}}} \;,
\label{delta-np1}
\ee
where $\t={\bf p}+\ro$, $\s={\bf q}+\ro$ and  ${\bf p}$, ${\bf q}$ and
\mbox{\boldmath$\nu$} are  local states respectively in the dominant  integral
weights  $P_+(n,l-p)$, $P_+(n,l)$ and $P_+(n,p)$. By the Feigin-Fuchs
construction
\cite{FeFu:82,DoFa:84},
$\delta_{_{\mbox{\boldmath$\nu$}}}$ is fixed by \cite{ChRa:89}
\be
\delta_{_{\mbox{\boldmath$\nu$}}}\;=\; {2p\mbox{\boldmath$\nu$}\cdot \ro -n
\mbox{\boldmath$\nu$}^2\over 2p(p+n)}\;.
\ee

In this paper we calculate the finite-size corrections to the eigenvalue
spectra of the transfer matrices of the fused JMO $A^{(1)}_{n-1}$ lattice
models.
Specifically, we generalize the analytic study introduced in
\cite{KlPe:92,KNS:93b} to
obtain the central charges and conformal weights of the fused JMO
$A^{(1)}_{n-1}$ lattice models.
The functional equations of the fused transfer matrices for the JMO
$A^{(1)}_{n-1}$ lattice model have been
given in \cite{BaRe:90}. In \cite{KuNa} the thermodynamic Bethe
ansatz-like equations (also called $y$-systems or inversion
identity hierarchies) of the model have been introduced and by solving
these functional equations Kuniba, Nakanishi and  Suzuki have extracted
the central charge (\ref{c-np}) from the finite-size corrections to
the eigenvalue spectra of the transfer matrices.
In fact the central charge and
conformal weights together appear in the finite-size corrections to
the eigenvalue spectra of the transfer matrices. To obtain the
conformal weights we have to consider the excited states of the
transfer matrices and the calculations  are more complicated
because of the more complicated analyticity for the states.
Therefore we  generalize the study done in
\cite{KNS:93,KNS:93b} for the fused JMO $A^{(1)}_{n-1}$ lattice models
to calculate some conformal weights.
In the scaling limit the fused JMO models have been shown~\cite{BaRe:90}
to yield the same  central charges (\ref{c-np})  as the $\W^n_p$ conformal
field
theories. Here we find the conformal weights
\be
\Delta_{t,s}\;=\;{n(n^2-1)\over 24}\;{[(l+n)t-(l+n-p)s]^2-
  p^2\over p(l+n-p)(l+n)}  +{2p\nu -n \nu^2\over 2p(p+n)} \;,
 \label{delta-np}
\ee where $1\;\le\;s\;  <\;\lf{l\over n-1}\rf$, $1\;\le\;t\;<\;\lf{l-p\over
n-1}\rf$ and
$\nu=(s-t)-\left\lfloor{s-t\over p}\right\rfloor p\;$, where
$\left\lfloor{x}\right\rfloor$ denotes the largest less than or equal
to x. Clearly, (\ref{delta-np1}) and (\ref{delta-np}) agree if
\be
\langle\t,\s\rangle&=&{n(n^2-1)\over 12}ts\;, \no \\
\t^2&=&{n(n^2-1)\over 12}t^2\;, \no \\
\s^2&=&{n(n^2-1)\over 12}s^2\;,  \label{case}\\
\mbox{\boldmath$\nu$}^2&=&{n(n^2-1)\over 12}\nu^2\;, \no \\
\langle\mbox{\boldmath$\nu$},\ro\rangle&=&{n(n^2-1)\over 12}\nu \no
\ee
This confirms that the underlying solvable statistical mechanics  models
corresponding to the conformal field theories with extended algebra
$\W^n_p$ are precisely the fused critical $A^{(1)}_{n-1}$ lattice models
of Jimbo, Miwa and Okado.

The paper is organized as follows. In the next subsection, we describe the
finite-size corrections to the eigenvalues of the row transfer matrices of
critical lattice models as predicted by conformal invariance. In section~2
we define
the  $A^{(1)}_{n-1}$ lattice models of Jimbo, Miwa and Okado. In particular, we
discuss the fusion rules satisfied by the adjacency matrices of these
models and the
corresponding functional equations of fused transfer matrices. These functional
equations can be converted into inversion identity hierarchies. These
inversion identity  hierarchies and their original functional
equations are the elementary relations needed to determine the finite-size
corrections to the eigenvalues of the row transfer matrices. Next we find the
asymptotic and bulk behavior of these inversion identity hierarchies, which
are the
important data for obtaining finite-size corrections. In section~3, we
convert these
functional equations into nonlinear integral equations which we solve
analytically.
After using some known~\cite{Kirillov:93} Rogers dilogarithm identities, we
obtain the
finite-size corrections, the central charges and conformal weights.
Finally, a brief
conclusion is given in section~4.

\subsection{Finite-size corrections}
Much work has been done on extracting the conformal spectra of exactly solvable
lattice models from finite-size corrections. These extensive calculations
\cite{DeWo:85,Wo:87,GeRi:87,DeKa:87,Ka:88,KiRe:87,ABB:87,BaRe:89,BNW:89,BaRe:90,KlPe:91,KlPe:92}
give very strong and direct evidence to support the
predictions of conformal and modular invariance.

According to the predictions of conformal and modular invariance, the partition
function of a two-dimensional lattice model on  a finite
$M\times N$ periodic lattice or torus, can be written for large $M$ and $N$ as
\be
Z_{M,N}=Tr {\T}^M\sim Z(q)e^{-MNf} \; .
\ee
Here $\T$ is the row transfer matrix, $f$ is the bulk free energy  and
$Z(q)$ is the universal finite-size partition function with modular  parameter
\be
q=e^{2\pi{\i} \tau} \; ,\h \tau={M\over N}\,e^{{\i}(\pi-hu)} \; .
\ee
Suppose that the eigenvalues of the row transfer matrix $\T$ of a periodic  row
of $N$ faces are given by
\be
\Lambda_n=e^{-E_n}
\ee
where $E_n$ are the corresponding energy levels. It then follows that
\be Z_{M,N}=\sum_n e^{-ME_n}.
\ee
Conformal
invariance now predicts~\cite{BCN:86,Affleck:86} that for large N the
energy levels
take the form
\be
E_0&\sim& Nf-{\pi c\over 6N}\sin (hu)\;, \\
E_n&\sim&E_0+{2\pi \over N}[{\it x}_n\sin
(hu)+{\i}s_n\cos(hu) ]
\ee
where $E_0$ is the ground-state energy,
\be
{\it x}_n=\Delta+\ol{\Delta}+k+\ol{k}\and
 s_n=\Delta-\ol{\Delta}+k-\ol{k}
\ee
are respectively the scaling dimensions and spins  of the various levels and
$k,\ol{k}$ are integers. The numbers
$c$ and $(\Delta,\ol{\Delta})$ are identified as the central charge and
conformal
weights of the primary operators of the underlying conformal field theory.

\section{The models and their fusion hierarchies}
\setcounter{equation}{0}

\subsection{Algebraic notations and Boltzmann weights}\label{notations}

\begin{figure}[t]
\begin{center}
\setlength{\unitlength}{0.006in}
\begin{picture}(633,603)(0,-10)
\drawline(378,453)(396,453)(396,435)(378,435)(378,453)
\drawline(378,453)(396,453)(396,435)(378,435)(378,453)
\drawline(396,453)(414,453)(414,435)(396,435)(396,453)
\drawline(396,453)(414,453)(414,435)(396,435)(396,453)
\drawline(414,453)(432,453)(432,435)(414,435)(414,453)
\drawline(414,453)(432,453)(432,435)(414,435)(414,453)
\drawline(432,453)(450,453)(450,435)(432,435)(432,453)
\drawline(432,453)(450,453)(450,435)(432,435)(432,453)
\drawline(378,435)(396,435)(396,417)(378,417)(378,435)
\drawline(378,435)(396,435)(396,417)(378,417)(378,435)
\drawline(189,456)(207,456)(207,438)(189,438)(189,456)
\drawline(189,456)(207,456)(207,438)(189,438)(189,456)
\drawline(171,456)(189,456)(189,438)(171,438)(171,456)
\drawline(171,456)(189,456)(189,438)(171,438)(171,456)
\drawline(153,456)(171,456)(171,438)(153,438)(153,456)
\drawline(153,456)(171,456)(171,438)(153,438)(153,456)
\drawline(276,336)(294,336)(294,318)(276,318)(276,336)
\drawline(276,336)(294,336)(294,318)(276,318)(276,336)
\drawline(258,336)(276,336)(276,318)(258,318)(258,336)
\drawline(258,336)(276,336)(276,318)(258,318)(258,336)
\drawline(258,318)(276,318)(276,300)(258,300)(258,318)
\drawline(258,318)(276,318)(276,300)(258,300)(258,318)
\drawline(294,336)(312,336)(312,318)(294,318)(294,336)
\drawline(294,336)(312,336)(312,318)(294,318)(294,336)
\drawline(450,336)(468,336)(468,318)(450,318)(450,336)
\drawline(450,336)(468,336)(468,318)(450,318)(450,336)
\drawline(450,318)(468,318)(468,300)(450,300)(450,318)
\drawline(450,318)(468,318)(468,300)(450,300)(450,318)
\drawline(468,318)(486,318)(486,300)(468,300)(468,318)
\drawline(468,318)(486,318)(486,300)(468,300)(468,318)
\drawline(468,336)(486,336)(486,318)(468,318)(468,336)
\drawline(468,336)(486,336)(486,318)(468,318)(468,336)
\drawline(486,336)(504,336)(504,318)(486,318)(486,336)
\drawline(486,336)(504,336)(504,318)(486,318)(486,336)
\drawline(504,336)(522,336)(522,318)(504,318)(504,336)
\drawline(504,336)(522,336)(522,318)(504,318)(504,336)
\drawline(126,333)(144,333)(144,315)(126,315)(126,333)
\drawline(126,333)(144,333)(144,315)(126,315)(126,333)
\drawline(108,333)(126,333)(126,315)(108,315)(108,333)
\drawline(108,333)(126,333)(126,315)(108,315)(108,333)
\drawline(204,207)(222,207)(222,189)(204,189)(204,207)
\drawline(204,207)(222,207)(222,189)(204,189)(204,207)
\drawline(186,207)(204,207)(204,189)(186,189)(186,207)
\drawline(186,207)(204,207)(204,189)(186,189)(186,207)
\drawline(186,189)(204,189)(204,171)(186,171)(186,189)
\drawline(186,189)(204,189)(204,171)(186,171)(186,189)
\drawline(354,207)(372,207)(372,189)(354,189)(354,207)
\drawline(354,207)(372,207)(372,189)(354,189)(354,207)
\drawline(336,207)(354,207)(354,189)(336,189)(336,207)
\drawline(336,207)(354,207)(354,189)(336,189)(336,207)
\drawline(318,207)(336,207)(336,189)(318,189)(318,207)
\drawline(318,207)(336,207)(336,189)(318,189)(318,207)
\drawline(318,189)(336,189)(336,171)(318,171)(318,189)
\drawline(318,189)(336,189)(336,171)(318,171)(318,189)
\drawline(336,189)(354,189)(354,171)(336,171)(336,189)
\drawline(336,189)(354,189)(354,171)(336,171)(336,189)
\drawline(525,189)(543,189)(543,171)(525,171)(525,189)
\drawline(525,189)(543,189)(543,171)(525,171)(525,189)
\drawline(525,207)(543,207)(543,189)(525,189)(525,207)
\drawline(525,207)(543,207)(543,189)(525,189)(525,207)
\drawline(543,207)(561,207)(561,189)(543,189)(543,207)
\drawline(543,207)(561,207)(561,189)(543,189)(543,207)
\drawline(561,207)(579,207)(579,189)(561,189)(561,207)
\drawline(561,207)(579,207)(579,189)(561,189)(561,207)
\drawline(579,207)(597,207)(597,189)(579,189)(579,207)
\drawline(579,207)(597,207)(597,189)(579,189)(579,207)
\drawline(543,189)(561,189)(561,171)(543,171)(543,189)
\drawline(543,189)(561,189)(561,171)(543,171)(543,189)
\drawline(561,189)(579,189)(579,171)(561,171)(561,189)
\drawline(561,189)(579,189)(579,171)(561,171)(561,189)
\drawline(57,207)(75,207)(75,189)(57,189)(57,207)
\drawline(57,207)(75,207)(75,189)(57,189)(57,207)
\drawline(144,39)(162,39)(162,21)(144,21)(144,39)
\drawline(144,39)(162,39)(162,21)(144,21)(144,39)
\drawline(144,21)(162,21)(162,3)(144,3)(144,21)
\drawline(144,21)(162,21)(162,3)(144,3)(144,21)
\drawline(282,36)(300,36)(300,18)(282,18)(282,36)
\drawline(282,36)(300,36)(300,18)(282,18)(282,36)
\drawline(300,36)(318,36)(318,18)(300,18)(300,36)
\drawline(300,36)(318,36)(318,18)(300,18)(300,36)
\drawline(300,18)(318,18)(318,0)(300,0)(300,18)
\drawline(300,18)(318,18)(318,0)(300,0)(300,18)
\drawline(282,18)(300,18)(300,0)(282,0)(282,18)
\drawline(282,18)(300,18)(300,0)(282,0)(282,18)
\drawline(438,39)(456,39)(456,21)(438,21)(438,39)
\drawline(438,39)(456,39)(456,21)(438,21)(438,39)
\drawline(456,39)(474,39)(474,21)(456,21)(456,39)
\drawline(456,39)(474,39)(474,21)(456,21)(456,39)
\drawline(420,39)(438,39)(438,21)(420,21)(420,39)
\drawline(420,39)(438,39)(438,21)(420,21)(420,39)
\drawline(420,21)(438,21)(438,3)(420,3)(420,21)
\drawline(420,21)(438,21)(438,3)(420,3)(420,21)
\drawline(438,21)(456,21)(456,3)(438,3)(438,21)
\drawline(438,21)(456,21)(456,3)(438,3)(438,21)
\drawline(456,21)(474,21)(474,3)(456,3)(456,21)
\drawline(456,21)(474,21)(474,3)(456,3)(456,21)
\drawline(561,36)(579,36)(579,18)(561,18)(561,36)
\drawline(561,36)(579,36)(579,18)(561,18)(561,36)
\drawline(579,36)(597,36)(597,18)(579,18)(579,36)
\drawline(579,36)(597,36)(597,18)(579,18)(579,36)
\drawline(597,36)(615,36)(615,18)(597,18)(597,36)
\drawline(597,36)(615,36)(615,18)(597,18)(597,36)
\drawline(615,36)(633,36)(633,18)(615,18)(615,36)
\drawline(615,36)(633,36)(633,18)(615,18)(615,36)
\drawline(561,18)(579,18)(579,0)(561,0)(561,18)
\drawline(561,18)(579,18)(579,0)(561,0)(561,18)
\drawline(579,18)(597,18)(597,0)(579,0)(579,18)
\drawline(579,18)(597,18)(597,0)(579,0)(579,18)
\drawline(597,18)(615,18)(615,0)(597,0)(597,18)
\drawline(597,18)(615,18)(615,0)(597,0)(597,18)
\drawline(615,18)(633,18)(633,0)(615,0)(615,18)
\drawline(615,18)(633,18)(633,0)(615,0)(615,18)
\drawline(312,588)(330,588)(330,570)(312,570)(312,588)
\drawline(312,588)(330,588)(330,570)(312,570)(312,588)
\drawline(294,588)(312,588)(312,570)(294,570)(294,588)
\drawline(294,588)(312,588)(312,570)(294,570)(294,588)
\drawline(276,588)(294,588)(294,570)(276,570)(276,588)
\drawline(276,588)(294,588)(294,570)(276,570)(276,588)
\drawline(258,588)(276,588)(276,570)(258,570)(258,588)
\drawline(258,588)(276,588)(276,570)(258,570)(258,588)
\put(72,174){\circle*{10}}\put(222,174){\circle*{10}}
\put(150,51){\circle*{10}}\put(297,51){\circle*{10}}
\put(369,174){\circle*{10}}\put(441,51){\circle*{10}}
\put(588,51){\circle*{10}}\put(516,174){\circle*{10}}
\put(441,303){\circle*{10}}\put(219,432){\circle*{10}}
\put(366,432){\circle*{10}}\put(4,51){\circle*{10}}
\put(145,303){\circle*{10}}\put(295,303){\circle*{10}}
\put(291,558){\circle*{10}}\put(93,102){\small$\hat 1$}
\put(16,102){\small$\hat 0$}\put(69,30){\small$\hat 2$}
\put(-6,20){\large$\phi$}
\thicklines
\drawline(3,51)(588,51)(511.25,180.5)(434.5,306)
\drawline(434.5,306)(362.75,433.5)(291,561)
\drawline(291,561)(147,306)(75,178.5)(3,51)
\drawline(297,51)(144.1,303)(441,303)(297,51)

\drawline(516,174)(444,51)
\drawline(72,174)(516,174)
\drawline(366,432)(150,51)(72,174)
\drawline(441,51)(219,432)(366,432)
\drawline(218.000,53.000)(210.000,51.000)(218.000,49.000)
\drawline(368.000,53.000)(360.000,51.000)(368.000,49.000)
\drawline(521.000,53.000)(513.000,51.000)(521.000,49.000)
\drawline(71.000,53.000)(63.000,51.000)(71.000,49.000)

\drawline(39.925,111.019)(42.000,119.000)(36.413,112.935)
\drawline(111.869,108.455)(117.000,102.000)(115.510,110.111)
\drawline(183.925,106.019)(186.000,114.000)(180.413,107.935)
\drawline(256.869,114.455)(262.000,108.000)(260.510,116.111)
\drawline(330.925,106.019)(333.000,114.000)(327.413,107.935)
\drawline(405.869,108.455)(411.000,102.000)(409.510,110.111)
\drawline(478.425,106.019)(480.500,114.000)(474.913,107.935)
\drawline(550.869,111.455)(556.000,105.000)(554.510,113.111)

\drawline(143.000,176.000)(135.000,174.000)(143.000,172.000)
\drawline(290.000,176.000)(282.000,174.000)(290.000,172.000)
\drawline(443.000,176.000)(435.000,174.000)(443.000,172.000)

\drawline(106.925,231.019)(109.000,239.000)(103.413,232.935)
\drawline(183.869,234.455)(189.000,228.000)(187.510,236.111)
\drawline(252.925,229.019)(255.000,237.000)(249.413,230.935)
\drawline(330.869,237.455)(336.000,231.000)(334.510,239.111)
\drawline(402.925,231.019)(405.000,239.000)(399.413,232.935)
\drawline(479.0,231.455)(484.100,225.000)(482.610,233.111)

\drawline(215.000,305.000)(207.000,303.000)(215.000,301.000)
\drawline(371.000,305.000)(363.000,303.000)(371.000,301.000)

\drawline(177.925,357.019)(180.000,365.000)(174.413,358.935)
\drawline(255.869,366.455)(261.000,360.000)(259.510,368.111)
\drawline(325.925,357.019)(328.000,365.000)(322.413,358.935)
\drawline(402.969,360.455)(408.100,354.000)(406.610,362.111)

\drawline(290.000,434.000)(282.000,432.000)(290.000,430.000)

\drawline(252.925,489.019)(255.000,497.000)(249.413,490.935)
\drawline(327.069,495.455)(332.200,489.000)(330.710,497.111)
\end{picture}
\end{center}
\caption{\label{fig1}The set of local states $P_+(3,4)$ ($n=3$, $l=4$).
         Each state is associated with a corresponding Young diagram and
         the admissible pairs of the states $(a,b)$ are represented by
         arrows from $a$ to $b$.}
\end{figure}
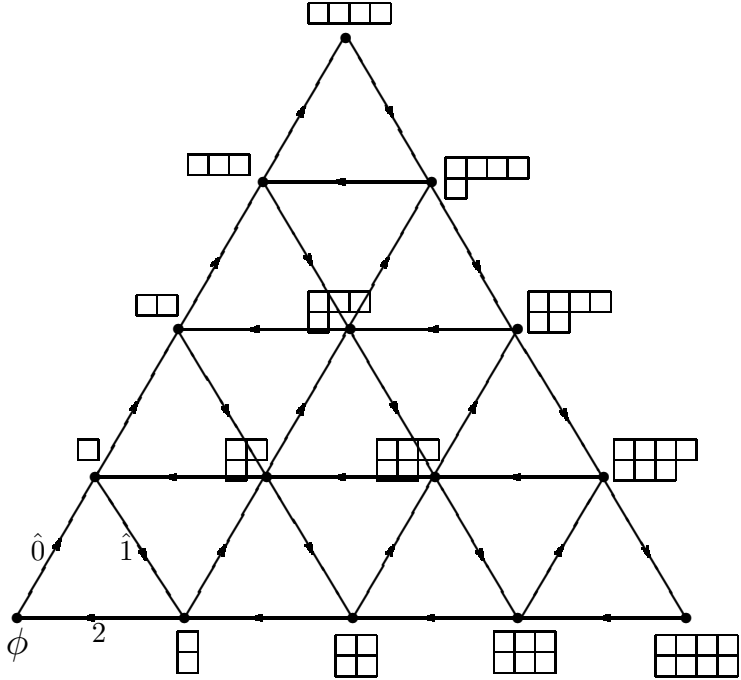

\noindent A vector $a$ represents a level $l$ dominant integral weight of
$A^{(1)}_{n-1}$ if
\be a=\sum_{\mu=0}^{n-1}a_\mu\Lambda_\mu, \h a_\mu\in \Z_+
\ee and $\sum_{\mu=0}^{n-1}a_\mu=l$, where $\Z_+$ is a set of all non-negative
integers and $\Lambda_\mu$ with $\mu=0,1,\cdots,n-1$ are  the fundamental
weights of
$A^{(1)}_{n-1}$ with $\Lambda_n=\Lambda_0$.  Fix an integer $l\ge 1$, and
denote by
$P_+(n,l)$ the set of dominant weights.  Then an element of $P_+(n,l)$ is
called a
local state.

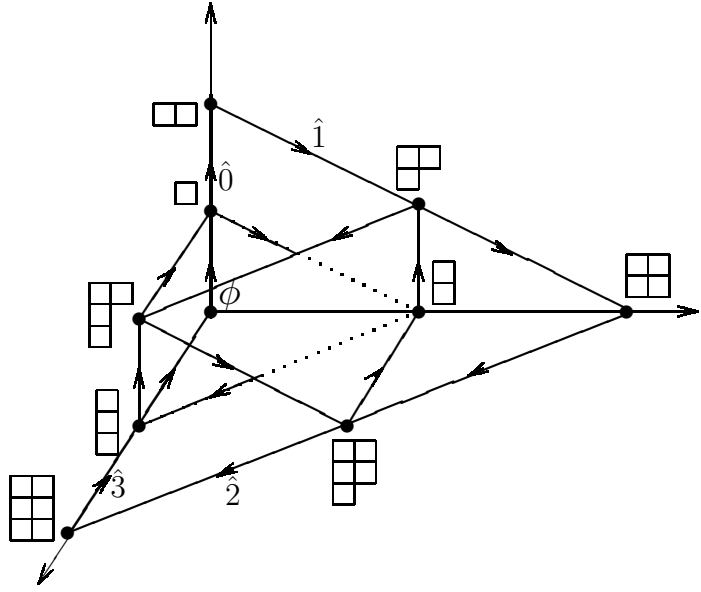
\begin{figure}[t]
\begin{center}
\setlength{\unitlength}{0.0125in}
\setlength{\unitlength}{0.0125in}
\begin{picture}(288,258)(0,-10)
\thicklines
\drawline(84,205)(84,114)(24,20)
\put(54,66){\circle*{6}}
\put(24,21){\circle*{6}}
\put(141,66){\circle*{6}}
\put(54,111){\circle*{6}}
\put(84,156){\circle*{6}}
\put(84,201){\circle*{6}}
\put(171,159){\circle*{6}}
\put(171,114){\circle*{6}}
\put(258,114){\circle*{6}}
\put(84,114){\circle*{6}}
\dottedline{5}(84,156)(171,114)
\dottedline{5}(171,114)(54,66)
\drawline(84,201)(261,114)
\drawline(261,114)(24,21)
\drawline(54,66)(54,111)(84,156)
\drawline(54,111)(171,159)(171,114)
\drawline(54,111)(141,66)(171,114)
\drawline(54,66)(105,87)
\drawline(84,156)(120,138)
\drawline(93,48)(87,45)
\thinlines
\drawline(93.261,50.367)(87.000,45.000)(95.050,46.789)
\drawline(204,141)(210,138)
\drawline(201.950,139.789)(210.000,138.000)(203.739,143.367)
\drawline(120,183)(126,180)
\drawline(117.950,181.789)(126.000,180.000)(119.739,185.367)
\drawline(141,147)(135,144)
\drawline(141.261,149.367)(135.000,144.000)(143.050,145.789)
\drawline(90,81)(84,78)
\drawline(90.261,83.367)(84.000,78.000)(92.050,79.789)
\drawline(171,129)(171,135)
\drawline(173.000,127.000)(171.000,135.000)(169.000,127.000)
\drawline(54,84)(54,90)
\drawline(56.000,82.000)(54.000,90.000)(52.000,82.000)
\drawline(84,129)(84,135)
\drawline(86.000,127.000)(84.000,135.000)(82.000,127.000)
\drawline(84,171)(84,177)
\drawline(86.000,169.000)(84.000,177.000)(82.000,169.000)
\drawline(87,93)(93,90)
\drawline(84.950,91.789)(93.000,90.000)(86.739,95.367)
\drawline(102,147)(108,144)
\drawline(99.950,145.789)(108.000,144.000)(101.739,149.367)
\drawline(66,129)(69,132)
\drawline(64.757,124.929)(69.000,132.000)(61.929,127.757)
\drawline(153,87)(156,90)
\drawline(151.757,82.929)(156.000,90.000)(148.929,85.757)
\drawline(84,111)(84,243)
\drawline(86.000,235.000)(84.000,243.000)(82.000,235.000)
\drawline(84,114)(288,114)
\drawline(280.000,112.000)(288.000,114.000)(280.000,116.000)
\drawline(198,90)(192,87)
\drawline(198.261,92.367)(192.000,87.000)(200.050,88.789)
\drawline(84,114)(12,0)
\drawline(14.581,7.832)(12.000,0.000)(17.963,5.696)
\drawline(63,81)(69,90)
\drawline(66.226,82.234)(69.000,90.000)(62.898,84.453)
\drawline(36,39)(42,45)
\drawline(37.757,37.929)(42.000,45.000)(34.929,40.757)
\put(42,36){$\hat 3$}
\put(90,33){$\hat 2$}
\put(87,165){$\hat 0$}
\put(126,183){$\hat 1$}
\put(87,117){\large$\phi$}
\drawline(0,45)(9,45)(9,36)
	(0,36)(0,45)
\drawline(9,36)(9,27)(0,27)(0,36)
\drawline(9,27)(9,18)(0,18)(0,27)
\drawline(9,45)(18,45)(18,36)
	(9,36)(9,45)
\drawline(18,36)(18,27)(9,27)(9,36)
\drawline(18,27)(18,18)(9,18)(9,27)
\drawline(36,81)(45,81)(45,72)
	(36,72)(36,81)
\drawline(45,72)(45,63)(36,63)(36,72)
\drawline(45,63)(45,54)(36,54)(36,63)
\drawline(33,126)(42,126)(42,117)
	(33,117)(33,126)
\drawline(42,117)(42,108)(33,108)(33,117)
\drawline(42,108)(42,99)(33,99)(33,108)
\drawline(42,126)(51,126)(51,117)(42,117)
\drawline(69,168)(78,168)(78,159)
	(69,159)(69,168)
\drawline(258,138)(267,138)(267,129)
	(258,129)(258,138)
\drawline(267,138)(276,138)(276,129)(267,129)
\drawline(258,129)(267,129)(267,120)
	(258,120)(258,129)
\drawline(267,129)(276,129)(276,120)(267,120)
\drawline(135,60)(144,60)(144,51)
	(135,51)(135,60)
\drawline(144,60)(153,60)(153,51)(144,51)
\drawline(135,51)(144,51)(144,42)
	(135,42)(135,51)
\drawline(144,51)(153,51)(153,42)(144,42)
\drawline(144,42)(144,33)(135,33)(135,42)
\drawline(60,201)(69,201)(69,192)
	(60,192)(60,201)
\drawline(69,201)(78,201)(78,192)(69,192)
\drawline(162,183)(171,183)(171,174)
	(162,174)(162,183)
\drawline(171,174)(171,165)(162,165)(162,174)
\drawline(171,183)(180,183)(180,174)(171,174)
\drawline(177,135)(186,135)(186,126)
	(177,126)(177,135)
\drawline(186,126)(186,117)(177,117)(177,126)
\end{picture}
\end{center}
\caption{\label{fig2}The set of local states $P_+(4,2)$ ($n=4$, $l=2$).
         Each state is associated with a corresponding Young diagram and
         the admissible pairs of the states $(a,b)$ are represented by the
         arrows from $a$ to $b$.}
\end{figure}

An ordered pair of local states $(a,b)$, with $a,b\in P_+(n,l)$, is called
admissible
if
\be b=a+\hat\mu, \h \mu=0,1,\cdots,n-1\;
\ee
where $\hat\mu$ are the elementary vectors defined by
\be
\hat{\mu}=\Lambda_{\mu+1}-\Lambda_{\mu}\h \mu=0,1,\cdots,n-1 \;.
\ee For each such pair we place an arrow from $a$ to $b$, then all  local
states in
set $P_+(n,l)$ can be represented by an oriented graph. There is a one-to-one
correspondence between a state in $P_+(n,l)$ and a Young diagram
$(f_1,\cdots,f_{n-1})$ with $l=f_0\ge f_1\ge \cdots\ge f_{n-1}\ge f_n=0$
(see Figures~\ref{fig1} and \ref{fig2})
given by
$$ a=\sum_{\mu=0}^{n-1}(f_\mu-f_{\mu+1})\Lambda_\mu. $$
It is usual to omit the columns of length
$n$ in these Young diagrams.

The $A^{(1)}_{n-1}$ lattice models are described by assigning a Boltzmann
weight
$$ W\left(\left. \matrix{d&c\cr a&b\cr}\right|u\right)=\face abcdu
$$ to each configuration $a,b,c,d\in P_+(n,l)$ of four sites surrounding  a
face. The face weights are nonzero only if
$(d,a),(a,b),(d,c),(c,b)$ are all admissible. The nonzero weights are given
by~\cite{JMO:87}
\be W\left(\left. \matrix{a           &a+\hat\alpha    \cr
                      a+\hat\mu   &a+\hat\mu+\hat\nu \cr}\right|u\right)
=\cases{\displaystyle{ {h(\lambda+u) \over h(\lambda)}},
                         &if $0\leq\alpha=\mu=\nu\leq n-1$;\cr
\noalign{\vskip9pt}
\displaystyle{{h(\lambda a^{\mu\nu}-u)\over h(\lambda a^{\mu\nu})}},
                         &if $0\leq\alpha=\mu\not=\nu\leq n-1$;\cr
\noalign{\vskip9pt}
\displaystyle{ {h(u) \over h(\lambda)}{h(\lambda
              a^{\mu\nu}+\lambda) \over h(\lambda a^{\mu\nu})}},
                        &if $0\le\alpha=\nu\not=\mu\le n-1$,\cr}
\ee
where $a$ is a local state and $a^{\mu\nu}$ is given by the inner-product
$\langle a+\rho,\hat \mu-\hat \nu\rangle$. Here $u$ is the spectral parameter,
$\lambda={\pi\over n+l}$,
$\rho$ is the  sum over all fundamental weights $\Lambda_\mu$ and $h(u)$ is the
elliptic theta function
\be
h(u)=\sin(u)\prod^{\infty}_{j=1}[(1-2 q^{2j}\cos 2u+q^{4j})(1-q^{2j})].
\ee
In this paper we will only
consider the critical case when the elliptic nome vanishes $q=0$ and the theta
function reduces to the trigonometric function
\be h(u)=\sin(u).
\ee

\subsection{su($n$) fusion rules}
The oriented graphs, such as that shown in Figure~\ref{fig1} and \ref{fig2},
describe all  admissible pairs $(a,b)$ in $P_+(n,l)$. The
corresponding adjacency matrix $A$  with elements
\be A_{a,b} =\cases{ 1      &\mbox{$(a,b)$ admissible}\cr 0
&\mbox{otherwise}}
\ee
satisfies certain $su(n)$ fusion rules which determine the admissible pairs
$(a,b)$ in $P_+(n,l)$ for the fused $A^{(1)}_{n-1}$ lattice models as we
will now
explain.

For fixed rank $n\ge 2$ and level $l$, the $su(n)$ fusion rules determine a
hierarchy
of  commuting adjacency matrices labeled by the representations
of $su(n)$. Set $\omega_1=(1,0,
\cdots,0)$ and consider the tensor product of two  irreducible
representations of
of $su(n)$ with Young tableaux
$f=(f_1,f_2,\cdots,f_{n-1})$ and $\omega_1$. Suppose the decomposition of
the tensor
product gives $s$ irreducible representations with Young tableaux
$f^k=(f^k_1,f^k_2,\cdots, f^k_{n-1})$ $k=1,2,\cdots s$ with $s<n$.  Then
the $su(n)$ fusion rule is expressed as
\be
A^{(f)}A=\sum_{k=1}^sA^{(f^k)}\; , \label{fusionrule-adj}
\ee
where $A^{(f)}=\1$ if $f=(0,0,\cdots,0)$ and $A^{(f)}=0$ unless $l\ge f_1\ge
\cdots\ge f_{n-1}\ge f_n=0$. In the oriented  adjacency graphs $s$ is just the
number of outgoing arrows originating from the local state with Young
tableau $f$ and
$f^1,f^2,\cdots f^s$ are the corresponding  adjacent local states to which the
outgoing arrows point (e.g. see Figure~\ref{fig1}). For example, suppose
that $n=3$
and $A^{(f)}$ is represented by the  Young diagram
$f=(f_1,f_2,\cdots,f_{n-1})$.
Then the fusion rule  (\ref{fusionrule-adj}) takes the form
\be
&&\setlength{\unitlength}{0.008in}%
\begin{picture}(279,70)(78,732)
\put(327,792){\line( 0,-1){ 30}}
\put(327,762){\line( 1, 0){ 30}}
\put(357,762){\line( 0, 1){ 30}}
\put(357,792){\line(-1, 0){ 30}}
\put(108,792){\line( 0,-1){ 30}}
\put(138,792){\line( 0,-1){ 30}}
\put(183,792){\line( 0,-1){ 30}}
\put(213,792){\line( 0,-1){ 30}}
\put(108,762){\line( 0,-1){ 30}}
\put(138,762){\line( 0,-1){ 30}}
\put(168,792){\line( 1, 0){ 60}}
\put(168,762){\line( 1, 0){ 60}}
\put(141,732){\line(-1, 0){ 63}}
\put( 78,732){\line( 0, 1){ 60}}
\put( 78,792){\line( 1, 0){ 63}}
\put( 78,762){\line( 1, 0){ 63}}
\put(183,792){\line( 0,-1){ 60}}
\put(183,732){\line(-1, 0){ 18}}
\put(252,792){\line( 1, 0){ 36}}
\put(288,792){\line( 0,-1){ 30}}
\put(288,762){\line(-1, 0){ 36}}
\put(258,792){\line( 0,-1){ 30}}
\put(300,768){$\otimes$}
\put(231,771){\small$p$}
\put(153,771){\small$q$}
\put(153,744){\small$q$}
\multiput(144,792)(4.,0.00000){6}{\makebox(0.4444,0.6667){\tenrm .}}
\multiput(144,762)(4.,0.00000){6}{\makebox(0.4444,0.6667){\tenrm .}}
\multiput(144,732)(4.,0.00000){6}{\makebox(0.4444,0.6667){\tenrm .}}
\multiput(231,762)(4.,0.00000){6}{\makebox(0.4444,0.6667){\tenrm .}}
\multiput(231,792)(4.,0.00000){6}{\makebox(0.4444,0.6667){\tenrm .}}
\end{picture} \no \\ \no \\
&&\setlength{\unitlength}{0.008in}%
\begin{picture}(441,60)(75,750)
\multiput(371,810)(4.00000,0.00000){7}{\makebox(0.4444,0.6667){\tenrm .}}
\multiput(371,780)(4.000,0.00000){7}{\makebox(0.4444,0.6667){\tenrm .}}
\multiput(371,750)(4.00000,0.00000){7}{\makebox(0.4444,0.6667){\tenrm .}}
\multiput(459,780)(4.00000,0.00000){5}{\makebox(0.4444,0.6667){\tenrm .}}
\multiput(459,810)(4.00000,0.00000){5}{\makebox(0.4444,0.6667){\tenrm .}}
\put(411,750){\line( 1, 0){ 30}}
\put(441,750){\line( 0, 1){ 30}}
\put(336,810){\line( 0,-1){ 30}}
\put(366,810){\line( 0,-1){ 30}}
\put(411,810){\line( 0,-1){ 30}}
\put(441,810){\line( 0,-1){ 30}}
\put(336,780){\line( 0,-1){ 30}}
\put(366,780){\line( 0,-1){ 30}}
\put(396,810){\line( 1, 0){ 60}}
\put(396,780){\line( 1, 0){ 60}}
\put(369,750){\line(-1, 0){ 63}}
\put(306,750){\line( 0, 1){ 60}}
\put(306,810){\line( 1, 0){ 63}}
\put(306,780){\line( 1, 0){ 63}}
\put(411,810){\line( 0,-1){ 60}}
\put(411,750){\line(-1, 0){ 18}}
\put(480,810){\line( 1, 0){ 36}}
\put(516,810){\line( 0,-1){ 30}}
\put(516,780){\line(-1, 0){ 36}}
\put(486,810){\line( 0,-1){ 30}}
\put(444,789){\small$p\!-\!1$}
\put(367,792){\small$q\!+\!1$}
\put(367,762){\small$q\!+\!1$}
\multiput(213,780)(4.0000,0.00000){8}{\makebox(0.4444,0.6667){\tenrm .}}
\multiput(213,810)(4.00000,0.00000){8}{\makebox(0.4444,0.6667){\tenrm .}}
\multiput(128,810)(4.00000,0.00000){7}{\makebox(0.4444,0.6667){\tenrm .}}
\multiput(128,780)(4.00000,0.00000){7}{\makebox(0.4444,0.6667){\tenrm .}}
\multiput(128,750)(4.00000,0.00000){7}{\makebox(0.4444,0.6667){\tenrm .}}
\put(126,810){\line(-1, 0){ 33}}
\put( 93,810){\line( 0,-1){ 60}}
\put( 93,750){\line( 1, 0){ 33}}
\put( 93,780){\line( 1, 0){ 33}}
\put( 93,810){\line( 0,-1){ 30}}
\put(123,810){\line( 0,-1){ 30}}
\put(168,810){\line( 0,-1){ 30}}
\put(198,810){\line( 0,-1){ 30}}
\put( 93,780){\line( 0,-1){ 30}}
\put(123,780){\line( 0,-1){ 30}}
\put(153,810){\line( 1, 0){ 60}}
\put(153,780){\line( 1, 0){ 60}}
\put(168,810){\line( 0,-1){ 60}}
\put(168,750){\line(-1, 0){ 18}}
\put(237,810){\line( 1, 0){ 36}}
\put(273,810){\line( 0,-1){ 30}}
\put(273,780){\line(-1, 0){ 36}}
\put(243,810){\line( 0,-1){ 30}}
\put( 75,777){\small$=$}
\put(215,789){\small$p$}
\put(127,789){\small$q\!-\!1$}
\put(127,762){\small$q\!-\!1$}
\put(284,777){$\oplus$}
\end{picture} \no \\ \no \\
&&\hs{0.3}\setlength{\unitlength}{0.008in}%
\begin{picture}(258,60)(75,750)
\put(303,810){\line( 1, 0){ 30}}
\put(333,810){\line( 0,-1){ 30}}
\put(333,780){\line(-1, 0){ 30}}
\put(123,810){\line( 0,-1){ 30}}
\put(153,810){\line( 0,-1){ 30}}
\put(198,810){\line( 0,-1){ 30}}
\put(228,810){\line( 0,-1){ 30}}
\put(123,780){\line( 0,-1){ 30}}
\put(153,780){\line( 0,-1){ 30}}
\put(183,810){\line( 1, 0){ 60}}
\put(183,780){\line( 1, 0){ 60}}
\put(156,750){\line(-1, 0){ 63}}
\put( 93,750){\line( 0, 1){ 60}}
\put( 93,810){\line( 1, 0){ 63}}
\put( 93,780){\line( 1, 0){ 63}}
\put(198,810){\line( 0,-1){ 60}}
\put(198,750){\line(-1, 0){ 18}}
\put(267,810){\line( 1, 0){ 36}}
\put(303,810){\line( 0,-1){ 30}}
\put(303,780){\line(-1, 0){ 36}}
\put(273,810){\line( 0,-1){ 30}}
\put( 75,780){$\oplus$}
\put(170,792){\small$q$}
\put(170,762){\small$q$}
\put(233,789){\small$p\!+\!1$}
\multiput(158,810)(4.00000,0.00000){7}{\makebox(0.4444,0.6667){\tenrm .}}
\multiput(158,780)(4.00000,0.00000){7}{\makebox(0.4444,0.6667){\tenrm .}}
\multiput(158,750)(4.00000,0.00000){7}{\makebox(0.4444,0.6667){\tenrm .}}
\multiput(243,780)(4.00000,0.00000){7}{\makebox(0.4444,0.6667){\tenrm .}}
\multiput(243,810)(4.00000,0.00000){7}{\makebox(0.4444,0.6667){\tenrm .}}
\end{picture} \no
\ee
where $p=f_1-f_2$ and $q=f_2$. There is a level-rank duality of the adjacency
matrices. Let $\ol{f}$ be the Young tableaux obtained from $f$ by
transposing the
diagram. Then the adjacency matrix $A^{(f)}$ of models with rank $n$ and
level $l$ is
the same as the adjacency matrix $A^{(\ol{f})}$ of models with rank $l+1$
and level
$n-1$.

Of particular interest here are weights $p\Lambda_a$ in $P_+(n,l)$ with
rectangular
Young tableaux $f=(f_1, f_2,\cdots,f_{n-1})$ where $f_{s\le a}=p$ and
$f_{s>a}=0$.
Let us write the corresponding adjacency matrix as
$A^{(a,p)}$. Then, from the fusion rule (\ref{fusionrule-adj}),  it follows
that
\be
A^{(a,p)}A^{(a,p)}=A^{(a,p-1)}A^{(a,p+1)}+A^{(a-1,p)}A^{(a+1,p)}\; ,
\label{func-adj}
\ee
where $a=1,2,\cdots,n-1$ and $p=1,2,\cdots,l$. Here $A^{(a,p)}=\1$ if
$a=0$ or $p=0$ and the closure condition is that $A^{(a,p)}=0$ if $p<0$ or
$p>l$ or
$a<0$ or $a>n$. In the next subsection~\ref{Function} we will see that a
similar
relation holds at the level of the transfer matrices of the fused models.

In general, the admissible states of adjacent sites of the fused models are
given by
the $su(n)$ fusion rule. In the appendix we give the explicit solution of
the fusion
rules for $n=4$ and $l=2$. In contrast to fusing the ABF  models, the
elements of $A^{(f^k)}$ for $n>2$ can in general be nonnegative  integers
greater
than one, for example, $A^{(3,1)}$ for $n=3$ and $l=5$. In such cases we have
to
distinguish the oriented edges between the adjacency states $(a,b)$ by bond
variables $\alpha=1,2,\ldots, A^{(f)}_{a,b}$.

\subsection{Functional equations}
\label{Function} A class of fused $A^{(1)}_{n-1}$ lattice models with adjacency
matrices $A^{(f)}$ can be constructed by the fusion procedure
\cite{JKMO:88}. Let
us consider the fused models obtained by fusing $(a\ p)\times(b\ q)$
elementary blocks
corresponding to the representations  $p\Lambda_a$ and $q\Lambda_b$. The
fused face
weights of these models vanish, that is,
 $$\WF {a}u{b}{c}{d}{\mu}{\beta}{\nu}{\alpha}\;=
\setlength{\unitlength}{0.0125in}%
\begin{picture}(48,40)(60,755)
\thicklines
\put( 75,780){\line( 0,-1){ 36}}
\put( 75,744){\line( 1, 0){ 30}}
\put(105,744){\line( 0, 1){ 36}}
\put(105,780){\line(-1, 0){ 30}}
\put(87,759){\sc$u$}
\put(64.8,757.5){\small$\alpha$}
\put(108,757.5){\small$\beta$}
\put(108,778){\sc$c$}
\put(108,738){\sc$b$}
\put( 66,738){\sc$a$}
\put(85.8,734){\small$\mu$}
\put(66,778){\sc$d$}
\put(85.8,783){\small$\nu$}
\end{picture}  \h=0 \h  $$ unless $A^{(a,p)}_{d,a}\ne 0$,
$A^{(a,p)}_{c,b}\ne 0$,
$A^{(b,q)}_{a,b}\ne 0$ and $A^{(b,q)}_{d,c}\ne 0$ and $1\le\alpha\le
A^{(a,p)}_{d,a}$,
$1\le\beta\le A^{(a,p)}_{c,b}$, $1\le\mu\le A^{(b,q)}_{a,b}$ and
$1\le\nu\le A^{(b,q)}_{d,c}$.  Suppose that $\mbox{\boldmath
$a$}(\mbox{\boldmath
$\alpha$})$ and
$\mbox{\boldmath $b$}(\mbox{\boldmath $\beta$})$ are  allowed state (bond)
configurations  of two consecutive rows of a lattice with $N$  columns and
periodic  boundary conditions. The elements of the fused row transfer matrices
${\bf T}^{(b,q)}_{(a,p)}(u)$ are given \vspace*{-1.7cm}by
\begin{eqnarray}
\langle\mbox{\boldmath $a,\alpha$}|{\bf T}^{(b,q)}_{(a,p)}(u)|\mbox{\boldmath
$b,\beta$}
 \rangle =
\prod_{j=1}^N\!\sum_{\{\eta_j\}} \WF {a_j}u{b_j}{b_{j\+1}}{a_{j\+1}}{\mh\eta_j
        \mh}{\beta_j}{\mh\eta_{j\+1}\mh}{\alpha_j}=
\setlength{\unitlength}{0.0115in}%
\begin{picture}(48,90)(60,760)
\thicklines
\put( 75,780){\line( 0,-1){ 36}}
\put( 75,744){\line( 1, 0){ 30}}
\put(105,744){\line( 0, 1){ 36}}
\put(105,780){\line(-1, 0){ 30}}
\multiput(105,810)(0.00000,-7.82609){12}{\line( 0,-1){  3.913}}
\multiput( 75,810)(0.00000,-8.28571){11}{\line( 0,-1){  4.143}}
\put( 87,759){\sc$u$}
\put( 62,756){\small$\alpha_j$}
\put(108,756){\small$\beta_j$}
\put(108,774){\sc$b_{j\+1}$}
\put(108,738){\sc $b_j$}
\put( 62,738){\sc$a_j$}
\put( 57,774){\sc $a_{\!j\+1}$}
\end{picture}
\ee\vspace*{0.3cm}\newline where $a_{N+1}=a_1$, $b_{N+1}=b_1$,
$\eta_{N+1}=\eta_1$ and $N=0\;{\rm mod}\;n$. Since the fused face weights
satisfy the
Yang-Baxter relation, we obtain a
hierarchy of commuting families of  transfer matrices. Specifically, if
$(a,p)$ are
held fixed
\begin{equation} [{\bf T}^{(b,q)}(u),{\bf T}^{(b',q')}(v)] = 0,
\label{rowcommute}
\end{equation}  where we have suppressed the subscripts $\T^{(b,q)}(u)
=\T^{(b,q)}_{(a,p)}(u)$. These transfer matrices satisfy a group of functional
equations which are expressed in determinant form in \cite{BaRe:90}. From this
determinant form the following useful functional equations can be extracted
\cite{KuNa},
\be &&\hs{2}\T^{(b,q)}(u)\T^{(b,q)}(u-\lambda)\;=\; \no\\
&&\T^{(b,q+1)}(u)\T^{(b,q-1)}(u-\lambda)+
\T^{(b+1,q)}(u)\T^{(b-1,q)}(u-\lambda)\;  \label{func-T}
\ee
These functional equations can be established using the fusion
procedure~\cite{JKMO:88} and they are the Baxterization  of the $su(n)$
fusion rule
(\ref{func-adj}) for the adjacency matrices.

The functional equations are easily converted into  the so called thermodynamic
Bethe ansatz-like (TBA-like) equations in the sense of \cite{KuNa,ZhPe:94}.
Thus
inserting
\be
\t^{(b,q)}(u):={\T^{(b,q+1)}(u)\T^{(b,q-1)}(u-\lambda)\over
\T^{(b+1,q)}(u)\T^{(b-1,q)}(u-\lambda)}     \label{def-t}
\ee into (\ref{func-T}) yields the  TBA-like hierarchy
\be
\t^{(b,q)}(u)\t^{(b,q)}(u-\lambda)=
{[\1+\t^{(b,q+1)}(u)][\1+\t^{(b,q-1)}(u-\lambda)]\over
[\1+(\t^{(b+1,q)}(u))^{-1}][\1+(\t^{(b-1,q)}(u-\lambda))^{-1}]}
\label{TBA}\ee
The closure condition becomes
\be
\t^{(b,0)}(u)=\t^{(b,l)}(u)=0\;, \h b=1,2,\cdots,n-1\;.
\ee By definition we have the following symmetry. Applying the replacement
\be
\t^{(b,q)}(u)\to \tt^{(b,q)}(u)={1\over \t^{(q,b)}(u)}
\ee to the TBA-like hierarchy, we have
\be
\tt^{(b,q)}(u)\tt^{(b,q)}(u-\lambda)=
{[\1+\tt^{(b,q+1)}(u)][\1+\tt^{(b,q-1)}(u-\lambda)]\over
[\1+(\tt^{(b+1,q)}(u))^{-1}][\1+(\tt^{(b-1,q)}(u-\lambda))^{-1}]}
\label{TBA1}
\ee with the closure condition
\be
\tt^{(b,0)}(u)=\tt^{(b,n)}(u)=0\;, \h b=1,2,\cdots,l-1\;. \label{tt-cl}
\ee
This symmetry is just level-rank duality \cite{SaAt:91,KuNa:89}.

In terms of the $\t$ matrices, the functional equations (\ref{func-T}) can
be rewritten as
\be
\T^{(a,p)}(u)\T^{(a,p)}(u-\lambda)=
\T^{(a-1,p)}(u-\lambda)\T^{(a+1,p)}(u)\(1+\t^{(a,p)}(u)\)\label{T-t1}
\ee It is obvious to see that we need to solve the TBA-like equations
(\ref{TBA})
first to get the solutions for the transfer matrices
$\T^{(a,p)}(u)$.

\subsection{Asymptotics of $t$ }
\label{asymptotic} The functional equations (\ref{TBA}) are easily solved in
the
braid limit $u\to\pm{\i}\infty$.  The asymptotics
$t^{(b,q)}_\infty=t^{(b,q)}(\pm{\i}\infty)\;$ satisfy
\be t^{(b,q)}_\infty t^{(b,q)}_\infty=
{[1+t^{(b,q+1)}_\infty][1+t^{(b,q-1)}_\infty]\over
[1+(t^{(b+1,q)}_\infty)^{-1}][1+(t^{(b-1,q)}_\infty)^{-1}]} \label{braid-t}
\ee with the closure condition
\be t^{(b,0)}_\infty=t^{(b,l)}_\infty=0\;, \h b=1,2,\cdots,n-1\;.
\label{braid-t-cl}
\ee Let us write $t^{(1,1)}_\infty$ as
\be t^{(1,1)}_\infty={\sin[(n+1)\theta]\over\sin[(n-1)\theta]}
\ee with $\theta$ to be determined. Using (\ref{braid-t}) as  a recursion
relation
for $t^{(b,q)}_\infty$ we then derive
\be t^{(b,q)}_\infty={\sin[(n+q)\theta]\sin[q\theta]\over
\sin[(n-b)\theta]\sin[b\theta]} \label{aspm-t}
\ee for $b=1,2,\cdots,n-1$ and $q=1,2,\cdots,l-1$. The closure condition
(\ref{braid-t-cl}) imposes the quantization
\be
\theta={m_j\pi\over h} \;,\h m_j=1,2,\cdots,\lf{l\over n-1}\rf ,
\label{chop}\ee
where $h=n+l$. Mathematically, the exponent $m_j$ can go from
$1$ to $h-1$. But $t^{(b,q)}_\infty$ with $m_j>\lf{l\over n-1}\rf$
is on longer the right solutions of the models. This can be checked
by looking the adjacency matrices or the transfer matrices for small
size.

We would like to mention here that the solutions given by (\ref{aspm-t})
are not the general ones of the equations (\ref{braid-t}). However,
the general solutions  involve  complex eigenvalues, which together
with the real eigenvalues (\ref{aspm-t}) give all of the low-lying
excited states.


\subsection{Zeros and poles of the transfer matrices}
\label{sec:zeros} To find the finite-size correction we need to solve the TBA
equations  (\ref{TBA}) or (\ref{TBA1}) in certain analytic domains. Inside
these
analyticity strips the ground state eigenvalues
$T^{(b,q)}(u)=T^{(b,q)}_{(a,p)}(u)$ should not  possess any zero except
those which
come from the parameterization   of  Boltzmann weights. They are of order
$N$ and
their locations are  independent of the eigenvalue under consideration.
Using self explanatory notation, the location of these zeros are as follows:
\be
&&\bigcup_{j=0}^{a-1}\bigcup_{l=0}^{p-1}\bigcup_{i=0}^{b-1}\bigcup_{k=0}^{q-2}
 \{(i\!+\!k\!-\!j\!-\!l)\lambda\} \bigcup_{j=0}^{a-1}\bigcup_{l=0}^{p-1}
\bigcup_{i=0}^{b-2}\{(q\!+\!i\!-\!j\!-\!l)\lambda\}\;,
       \hs{0.3} q\ge p\;\and\;b\ge a\h\h\\
&&\bigcup_{j=0}^{a-1}\bigcup_{l=0}^{p-1}\;\bigcup_{i=0}^{b-1}\bigcup_{k=0}^{q-2}
 \{(i\!+\!k\!-\!j\!-\!l)\lambda\} \bigcup_{j=1}^{a-1}\bigcup_{l=0}^{p-1}
 \bigcup_{i=0}^{b-1}\{(q\!+\!i\!-\!j\!-\!l)\lambda\}\;,
       \hs{0.3} q\ge p\;\and\;b\le a\h\h\\
&&\bigcup_{j=0}^{a-1}\bigcup_{l=1}^{p-1}\;\bigcup_{i=0}^{b-1}\bigcup_{k=0}^{q-1}
   \{(i\!+\!k\!-\!j\!-\!l)\lambda\} \bigcup_{i=1}^{b-1}\bigcup_{k=0}^{q-1}
\bigcup_{j=0}^{a-1}\{(k\!+\!i\!-\!j)\lambda\}\;,
       \hs{0.7}  q\le p\;\and\;b\ge a\h\h\\
&&\bigcup_{j=0}^{a-1}\bigcup_{l=1}^{p-1}\;\bigcup_{i=0}^{b-1}\bigcup_{k=0}^{q-1}
   \{(i\!+\!k\!-\!j\!-\!l)\lambda\}\bigcup_{i=0}^{b-1}\bigcup_{k=0}^{q-1}
\bigcup_{j=0}^{a-2}\{(k\!+\!i\!-\!j)\lambda\}\;,
       \hs{0.7} q\le p\;\and\;b\le a\h\h
\ee
These can be seen from the fusion procedure of the models \cite{JKMO:88}.
To locate
the zeros and poles of $t^{(b,q)}$ we use the definition (\ref{def-t})
and the zeros of $T^{(b,q)}(u)$.  We distinguish several cases as follows:
\be
&{\rm (I)}& 1\le q\le p-1 \;\and\; b=a  \hs{4} \no\\ &&\mbox{zeros: }
\;\mbox{$\emptyset$}   \no\\  &&\mbox{poles: }\bigcup_{l=1}^{q}\{l\lambda\}
\\ &{\rm (II)}&
q=p \;\and\; 1\le b\le a-1  \hs{0.5} \no\\ &&\mbox{zeros: }
\bigcup_{j=1}^{b}\{(j-a)\lambda\}   \no\\  &&\mbox{poles:
}\;\mbox{$\emptyset$}  \\ &{\rm
(III)}& q=p \;\and\; b=a     \no\\ &&\mbox{zeros:
}\bigcup_{j=0}^{a-1}\{-j\lambda\}
\no\\ &&\mbox{poles: }\bigcup_{l=1}^{p}\;\{ \; l \lambda\; \}      \\ &{\rm
(IV)}&
q\ge p+1 \;\and\; b=a  \hs{0.5} \no\\ &&\mbox{zeros: } \;\mbox{$\emptyset$}
\no\\
&&\mbox{poles: }\bigcup_{l=1}^{p}\{(l+q-p)\lambda\} \\ &{\rm (V)}& q=p \;\and\;
b\ge a+1  \hs{0.5} \no\\ &&\mbox{zeros: } \bigcup_{j=1}^{a}\{(j-a)\lambda\}
\no\\
&&\mbox{poles: }\;\mbox{$\emptyset$}      \\ &{\rm (VI)}& q\not=p \;\and\;
b\not=a  \hs{0.5}
\no\\ &&\mbox{zeros: } \;\mbox{$\emptyset$}  \no\\  &&\mbox{poles:
}\;\mbox{$\emptyset$}
\ee This pattern of zeros and poles is divided according to the fusion level
$(a,p)$  in the vertical direction. We fix the vertical fusion level $(a,p)$
and
vary the fusion level $(b,q)$ in the horizontal direction.   The transfer
matrices
$\t^{(b,q)}$ are free of zeros for
$q\ne p$ and  free of poles for $b\ne a$.

\subsection{Bulk behavior and the largest eigenvalues}
\label{bulk} According to section~\ref{sec:zeros} the analyticity strip for
$t^{(a,p)}(u)$ contains a zero of order $N$ at $u=0$ and a poles  of order
$N$ at $u=\lambda$.  All other functions $t^{(b,q)}$ are  analytic and
non-zero in their analyticity strips $-\half\lambda\le  u\le\half\lambda$.
For large $N$ the leading bulk behavior to $t^{(b,q)}$ is given by
\be t^{(b,q)}_{\rm bulk}(u)=\left\{ \begin{array}{ll}
 \mbox{constant ,}                          & q\not=p\;\Or\;b\not=q\;,\\
 \mbox{constant}\[\tan({1\over 2}h u)\]^N\;,& q=p \and b=a \; .
\ea \right. \label{bulk-t}
\ee
The ansatz of bulk behavior matches the zero and pole distribution. The
constants are fixed by the TBA-like equations (\ref{TBA}) and can be calculated
similarly to the asymptotics of $t^{(b,q)}$.  Corresponding to the asymptotics
solutions (\ref{aspm-t}) we find that the bulk values
$t^{(b,q)}_{\rm bulk}$ for $1\le q\le p-1$ and $1\le b\le n-1$ are given by
\be
t^{(b,q)}_{\rm bulk}={\sin[(q+n)\sigma]\sin(q\sigma)\over
\sin(b\sigma)\sin[(n-b)\sigma]}
\label{bs1}
\ee with
\be
\sigma={m'_j\pi\over p+n} \h m'_j=1,2,\cdots,\lf{p\over n-1}\rf
\ee and for $p+1\le q\le h-n-1$ and $1\le b\le n-1$ by
\be t^{(b,q)}_{\rm bulk}={\sin[(q-p+n)\tau]\sin[(q-p)\tau]\over
\sin(b\tau)\sin[(n-b)\tau]}
\label{solution-2}  \label{solution2}\label{bs2}\ee with
\be
\tau={m^{''}_j\pi\over h-p} \h m^{''}_j=1,2,\cdots,\lf{l-p\over n-1}\rf.
\ee Here we see that $p=1,2,\cdots,h-2$. For the largest eigenvalue, the
appropriate choices are $\theta={\pi/ h}$,
$\sigma={\pi/( p+n)}$ and $\tau={\pi/( h-p)}$ in  (\ref{aspm-t}),
(\ref{bs1}) and (\ref{bs2}) respectively. Here for the same reason as
the braid limit (\ref{chop}) we restrict the range of
the exponents $m^{'}_j,m^{''}_j$.

\section{Finite-size corrections}
\setcounter{equation}{0}

The finite-size corrections for the eigenvalues
$T^{(a,p)}$ can be obtained by solving  the equation
\be T^{(a,p)}(u)T^{(a,p)}(u-\lambda)=
T^{(a-1,p)}(u-\lambda)T^{(a+1,p)}(u)\(1+t^{(a,p)}(u)\)\label{T-t}
\ee
Although the JMO $A^{(1)}_{n-1}$ models satisfy an
inversion relation, there is no crossing symmetry for the face
weights of these models so we cannot extract the bulk behavior for the
transfer matrices from the  inversion relation alone. Instead the finite-size
correction to
$T^{(a,p)}(u)$ is given by
\be T^{(a,p)}(u)=T^{(a,p)}_{\rm bulk}(u)T^{(a,p)}_{\rm finite}(u) \; .
\label{bulk-finite}\ee Inserting (\ref{bulk-finite}) into (\ref{T-t}) and
setting
the bulk behavior as
\be T^{(a,p)}_{\rm bulk}(u)T^{(a,p)}_{\rm bulk}(u-\lambda)=
T^{(a-1,p)}(u-\lambda)T^{(a+1,p)}(u) \label{T-bulk}
\ee we find
\be T^{(a,p)}_{\rm finite}(u)T^{(a,p)}_{\rm finite}(u-\lambda)
=1+t^{(a,p)}(u) \;
.\label{finite}
\ee We can check that (\ref{T-bulk}) is correct for the case of $n=2$ or the
$(l+1)$-state ABF models in \cite{KlPe:92}. The finite-size  corrections for
$T^{(a,p)}(u)$, therefore, are represented by  the inversion identity hierarchy
$t^{(a,p)}(u)$. In the  analytical treatment of (\ref{finite}) and
(\ref{TBA}), we will see that the only inputs for the finite-size
corrections in the
scaling limit are the asymptotics and bulk behavior.

\subsection{Nonlinear equations for finite-size corrections}

It is useful to introduce functions of a real variable by restricting the
eigenvalue functions to certain lines in the complex plane,
\be {\U}^{(b,q)}(x)&:=&1+{\it \a}^{(b,q)}(x) \;,  \vs{0.3}  \\
{\a}^{(b,q)}(x)&:=&t^{(b,q)}\({n{\i}\over 2h}x-{a-b+p-q\over 2}\lambda\)\;, \\
{\b}^{(b,q)}(x)&:=&T^{(b,q)}_{\rm finite}\({n{\i}\over 2h}x-{a-b+p-q+1\over
2}\lambda\)\;.
\ee The functional relation (\ref{finite}) can then be rewritten in terms
of  the
new functions as
\be
\b^{(a,p)}(x-\pi{\i}/n)\b^{(a,p)}(x+\pi{\i}/n)={\U}^{(a,p)}(x)\; .
\ee For the ground state the functions $\U^{(b,q)}$ and $\b^{(b,q)}$ are {\sl
analytic, non-zero} in $-2\pi/n<x<2\pi/n$ and  possess {\sl constant}
asymptotics
for $ \re x\to \pm \infty$ (ANZC). Taking the logarithmic derivative of the
above
equation and introducing Fourier transforms
\be &&{\cal B}^{(a,p)}(k)={1\over 2\pi}\int_{-\infty}^{\infty}
dx\;[\ln\b^{(a,p)}(x)]'\;e^{-\i kx}
\;,\no\\ &&\h [\ln\b^{(a,p)}(x)]'=\int_{-\infty}^{\infty} dk\;{\cal
B}^{(a,p)}(k)
\;e^{\i kx} \;
\ee with analogous equations for $\U^{(a,p)}$ and its Fourier transform
$A^{(a,p)}$, we then find that
\be
{\cal B}^{(a,p)}(k)={A^{(a,p)}\!(k)\over e^{(\pi/n)k}+e^{-(\pi/n)k}}\;.
\ee Transforming back and defining the kernel $k(x)$
\be k(x):= {n\over 4\pi\cosh(nx/2)} \;,
\ee we are able to express $\b^{(a,p)}$ in terms of $\U^{(a,p)}$,
\be
\ln\b^{(a,p)}=k*\ln \U^{(a,p)} +C^{(a,p)}\;, \label{b}
\ee where $C^{(a,p)}$ are integration constants.  The convolution $f*g$ of two
functions $f$ and $g$ is defined by
\be (f*g):=\int^\infty_{-\infty}f(x-y)g(y)\;dy=
\int^\infty_{-\infty}g(x-y)f(y)\;dy\; .
\label{convolution}\ee  In the case of low-lying excitations we have to
take care of the zeros  in the analyticity strips so that the simple ANZC
properties
hold. The result  (\ref{b}) is still correct if we change the integration
path ${\cal
L}$ so that
$b^{(a,p)}(x)$ has an ANZC area and Cauchy's theorem can be applied as in
\cite{KlPe:92}.  The integration constants in (\ref{b}) can be evaluated
from the asymptotics of $\U^{(a,p)}$ and $\b^{(a,p)}$. In this limit
(\ref{b}) becomes
\be
\ln\b^{(a,p)}_\infty=\half\ln \U^{(a,p)}_\infty +C^{(a,p)}\;.
\ee It can be seen that the constants are dependent on the system size
$N$ and do not contribute to the $1\over N$ corrections.

The $\U^{(a,p)}$  is from the inversion identity hierarchy and  can be
solved from
(\ref{TBA}), which can be rewritten in terms  of $\a^{(b,q)}$ as
\be {\a^{(b,q)}(x-\pi{\i}/n)\a^{(b,q)}(x+\pi{\i}/n)\over
\a^{(b-1,q)}(x)\a^{(b+1,q)}(x)}
={\U^{(b,q-1)}(x)\U^{(b,q+1)}(x)\over{{\U}}^{(b-1,q)}(x){{\U}}^{(b+1,q)}(x)}\;
{}.
\ee We introduce finite-size correction terms $l^{(b,q)}(x)$ by writing
$\a^{(b,q)}(x)$ as
\be
\a^{(b,q)}(x)=\left\{\begin{array}{ll}
 l^{(b,q)}(x)\; , & q\not=p \;\Or \;b\not=a\\
\tanh^N(nx/4)l^{(a,p)}(x)\; , & q=p \and b=a
\end{array}\right. \label{a-l}.
\ee For the ground state all the functions $l^{(b,q)}(x)$ are  ANZC in
$-\pi<x<\pi$. They satisfy the functional equations
\be {l^{(b,q)}(x-\pi{\i}/n)l^{(b,q)}(x+\pi{\i}/n)\over l^{(b-1,q)}(x)
l^{(b+1,q)}(x)}=
{\U^{(b,q-1)}(x)\U^{(b,q+1)}(x)\over{{\U}}^{(b-1,q)}(x){{\U}}^{(b+1,q)}(x)}\; .
\ee Again applying Fourier transforms to the logarithmic derivative of the
equations
the Fourier transform $L^{(b,q)}(k)$ to $l^{(b,q)}(x)$ satisfies
\be &&L^{(b-1,q)}(k)-2\cosh\biggr({\pi k\over
n}\biggr)L^{(b,q)}(k)+L^{(b+1,q)}(k)\no \\
&&\;=\;A^{(b-1,q)}+A^{(b+1,q)}-A^{(b,q+1)}-A^{(b,q-1)}\; .
\ee For fixed $q$ we have the closure conditions
$L^{(0,q)}=L^{(n,q)}=A^{(0,q)}=A^{(n,q)}=0$. This set of $n-1$ linear equations
can be rewritten in matrix form as
\begin{equation}
\(Adj+K_0\)\cdot L^q(k)=Adj\cdot A^q -A^{q+1}-A^{q-1} \label{LA}
\end{equation} with
\begin{equation}
 K_0=-2\cosh\biggr({\pi k\over n}\biggr)             \; ,
\end{equation} where $L^q(k)$ is $(n-1)\times 1$ matrix with the elements
$L^{(b,q)}$
and the $(n-1)\times 1$ matrix $A^q(k)$ has the elements $A^{(b,q)}$. The
matrix
$Adj$ is the same as the adjacency matrix of the classical $A_{n-1}$ Dynkin
diagram.
The equations (\ref{LA}) are solvable. Similar equations in fact appear in
\cite{KlPe:92} and these can be solved using a similar method. Thus, after
integrating the
equations, we obtain the set of nonlinear integral equations
\be l\a^{q}=l\e^{q}+K*l\U^{q}-{\hat K}*\(\;l\U^{q+1}+l\U^{q-1}\;\) +D^{q}\; ,
\label{a}
\ee where the entries of the $(n-1)\times 1$ matrices $l\a^{q}$, $lU^{q}$ and
$l\e^{q}$
are respectively given by $\ln\a^{(b,q)}$, $\ln\U^{(b,q)}$ and
\be
\ln\e^{(b,q)}(x):=\left\{ \begin{array}{ll}        \label{e} 0 \; , &
q\not=p \;\Or
\;b\not=a\\
\ln\tanh^N(nx/4)\; , & q=p \and b=a\;.\end{array}\right.
\ee
The $(n-1)\times(n-1)$ matrices $K$ and $\hat K$ are the  symmetric
matrices whose
entries in the upper right triangle are given by
\be
\hat{K}^j_l(x)={1\over 2\pi}\int_{-\infty}^\infty dk\;e^{\i kx}\biggr(
    {\sinh [(n-l)\pi k/n]\sinh[j\pi k/n]\over
      \sinh[\pi k/n]\sinh(\pi k) }\biggr)
\ee
\be K^j_l(x)=-{1\over \pi}\int_{-\infty}^{\infty}dk\;e^{\i kx}\Biggr(
  \coth\biggr({\pi k\over n}\biggr){\sinh [(n-l)\pi k/n]
        \sinh[j\pi k/n]\over \sinh(\pi k) }\Biggr).
\ee
The set of equations (\ref{a}) are also obtained for the excited states
after we take care of the extra zeros so that the ANZC properties hold in
the analyticity
strips.

\subsection{Scaling limit}

The finite-size corrections can be extracted from the nonlinear
integral equations (\ref{a}) and (\ref{b}).  The system size $N$ enters the
nonlinear equations (\ref{a})  through (\ref{e}). The function $\e^{(b,q)}$ has
three asymptotic  regimes with transitions  in scaling regimes when $x$ is
of the
order  of $-\ln N$ or $\ln N$. We suppose that $\a^{(b,q)}$ and
$\U^{(b,q)}$  scale
similarly and define:
\be &&e^{(b,q)}_{\pm}(x):=\lim_{N\to \infty}\e^{(b,q)}\(\pm{2\over n}(x+\ln
N)\) \;
,   \no \\ &&a^{(b,q)}_{\pm}(x):=\lim_{N\to \infty}\a^{(b,q)}\(\pm{2\over
n}(x+\ln
N)\) \; ,\\ &&A^{(b,q)}_{\pm}(x):=\lim_{N\to \infty}\U^{(b,q)}\(\pm{2\over
n}(x+\ln
N)\)=1+a^{(b,q)}_\pm(x)\;,
\label{scaling}\ee In this scaling limit, (\ref{a}) takes the form
\be
\la^{q}=\Le^{q}+K*\lA^{q}-{\hat K}*\(\lA^{q+1}+\lA^{q-1}\) +D^{q}\; ,
\label{a-L}
\ee where we suppress the subscripts $\pm$. The entries of the $(n-1)\times
1$ matrices
$\a^{q}$, $U^{q}$ and $\e^{q}$  are respectively given by
$\la^{(b,q)}$, $\lA^{(b,q)}$ and $\Le^{(b,q)}$:
\be
 \la^{(b,q)}(x)&:=&\ln a^{(b,q)}(x) \; , \\
 \lA^{(b,q)}(x)&:=&\ln A^{(b,q)}(x) \; ,  \\
 \Le^{(b,q)}(x)&:=&\left\{\begin{array}{ll}
    0\; , &q\not=p \;\Or \;b\not=a\;, \\
  -2 e^{-x}\; ,& q=p \and b=a\; , \end{array}\right.
\ee
where $q=1,2,\cdots,l-1$ and $b=1,2,\cdots,n-1$ and
\be
a^{(b,q)}(x)&=&0 \h \mbox{$b<1$, $b>n-1$, $q<1$ and $q>l-1$} \\
\lA^{(b,q)}(x)&=&0 \h \mbox{$b<1$, $b>n-1$, $q<1$ and $q>l-1$}
\ee

Now let us consider the scaling limit of $\b^{(a,p)}(x)$, which gives the
finite-size corrections to $\b^{(a,p)}(x)$. Disregarding the integration
constants
in (\ref{b}), and for fixed $x$, the finite-size corrections to
$\b^{(a,p)}(x)$ are given by the following expression of order $1/N$,
\be &&\ln\b^{(a,p)}(x):=\lim_{N\to \infty}\;(k*\ln\U^{(a,p)})(2x/n) \no \\
&&={n\over 4\pi}\lim_{N\to \infty}\;\int^\infty_{_{-\!\ln
N}}\!dy\({\ln\U^{(a,p)}(y+\ln N)\over
 \cosh[x-n(y-\ln N)/2]}+{\ln\U^{(a,p)}(-y-\ln N)\over
 \cosh[x+n(y+\ln N)/2]}\) +{\Got o\!}\({1\over N}\)  \no\\ &&={e^x\over
N\pi}\int^\infty_{-\infty} e^{-y}\lA^{(a,p)}_+(y)\;dy +{e^{-x}\over
N\pi}\int^\infty_{-\infty} e^{-y}\lA^{(a,p)}_-(y)\;dy +{\Got o\!}\({1\over N}\)
\no \\ &&={\cosh x\over N\pi}\int^\infty_{-\infty} e^{-y}
\(\lA^{(a,p)}_+(y)+\lA^{(a,p)}_-(y)\)\;dy\no\\ &&\h +{\sinh x\over
N\pi}\int^\infty_{-\infty}
e^{-y}\(\lA^{(a,p)}_+(y)-\lA^{(a,p)}_-(y)\)\;dy+{\Got
o\!}\({1\over N}\)
\;.\label{b-N}
\ee It can be seen from (\ref{a-L}) that for $q=p$ and $b=a$ the  integrals in
the above equation converge because the function
$\lA^{(a,p)}(y)=\ln\(1+\la^{(a,p)}(y)\)$ tends to zero faster  than any
exponential. Moreover, as in the case of the critical ABF RSOS models, we have
$\lA^p_+(y)-\lA^p_-(y)=0$.

The integrals in (\ref{b-N}) exist  and can be evaluated  explicitly with
the help of the dilogarithmic function
\be L(x)=-\int_0^x dy\;{\ln (1-y)\over y} +\half \ln x\ln (1-x) .
\ee To show this, multiplying the derivative of (\ref{a-l})  with
$\lA^{(b,q)}$,
and (\ref{a-l}) itself with $(\lA^{(b,q)})'$,  then taking the difference and
summing over $q$ and $b$, and finally integrating we derive
\be &&\h\sum_{q=1}^{l-1}\sum_{b=1}^{n-1}\int_{-\infty}^{\infty}[(\la^{(b,q)})'
\lA^{(b,q)}-\la^{(b,q)}(\lA^{(b,q)})']\; dx=\no\\
&&\sum_{q=1}^{l-1}\sum_{b=1}^{n-1}\int_{-\infty}^{\infty}
[(\Le^{(b,q)})'\lA^{(b,q)}-(\Le^{(b,q)}-D^{(b,q)})(\lA^{(b,q)})']
\; dx \; ,  \label{la-lA}
\ee where the contribution of the kernel cancels
due to the symmetry
\be k(-x)=k(x)\; . \label{kernel-sym}
\ee
Then using the nonlinear integral equations (\ref{a}) to simplify the
right-hand
side and integrating the left-hand side of (\ref{la-lA}), we can write
the finite-size corrections as
\be &&\hs{2} 2\int_{-\infty}^\infty e^{-y}\lA^{(a,p)}(y)\;dy= \no\\
&&-\sum_{q=1}^{l-1}\sum_{b=1}^{n-1}L\({1\over A^{(b,q)}}\)
\rule[-15pt]{0.2mm}{35pt}_{-\infty}^\infty+
\half\sum_{q=1}^{l-1}\sum_{b=1}^{n-1}D^{(b,q)}
\lA^{(b,q)}\;\rule[-15pt]{0.2mm}{35pt}_{-\infty}^\infty
 \label{integral}\ee
where the constants $D^{(b,q)}=0$ which can be shown
by  the asymptotics of the equations (\ref{a-L}).

It can be seen from (\ref{integral}) that the finite-size corrections of the
transfer matrices in the scaling limit depend only on  the solutions in the
limits
$x\to\pm\infty$, which are given by the asymptotics and  bulk behavior.

\subsection{The central charge and the conformal weights}
The following useful dilogarithm identity has been established by
Kirillov~\cite{Kirillov:93}. Consider the functions
\be
 y^{(b,q)}\!(j,n,r):={\sin[(q+n)\varphi]\sin[q\varphi]\over
          \sin[a\varphi]\sin[(n-a)\varphi]}\; ,
 \h 1\le b\le n-1,\h 1\le q\le r \label{su(n)2}
\ee with
\be
\varphi={(1+j)\pi\over n+r }\h 0\le j\le n+r-2\;.
\ee
It is obvious that they are the asymptotic solutions of the  functions
equations (\ref{TBA}) with $r=l$ or the bulk behavior of the functions
equations with $r=p$ and $r=l-p$.  Then the following dilogarithmic
function identity holds,
\be s(j,n,l)&:=&
\sum_{k=1}^{n-1}\sum_{m=1}^{r}L({1\over 1+y^{(k,m)}\!(j,n,r)})
      \label{dilogarithm-s}      \no\\
&=&L(1)\({(n^2-1)r\over n+r}-{n(n^2-1)j(j+2)\over n+r} \\
&&+\;6j\left\lfloor{n\over 2}\right\rfloor\left\lfloor{n+1\over 2}
    \right\rfloor+6\Z_+\) \;.   \no
\ee
Here the brackets $\left\lfloor x\right\rfloor$ denotes the greatest
integer less than or
equal to $x$.

The finite-size corrections (\ref{b-N}) are expressed in terms of the
dilogarithm function in (\ref{integral})
\be
\ln\b^{(a,p)}(x)=-{\cosh x\over N\pi}\sum_{q=1}^{l-1}
  \sum_{b=1}^{n-1}L\({1\over A^{(b,q)}}\)
  \rule[-15pt]{0.2mm}{35pt}_{-\infty}^\infty \;.
\ee
The inputs are the asymptotic and bulk solutions
obtained in sections~\ref{asymptotic} and \ref{bulk}. Thus
we have
\be
\ln\b^{(a,p)}(x)&=&{\cosh x\over N\pi}\left(
 \sum_{b=1}^{n-1}\sum_{q=1}^{l-1}L\({1\over A^{(b,q)}(-\infty)}\)
-\sum_{b=1}^{n-1}\sum_{q=1}^{l-1}L\({1\over A^{(b,q)}(\infty)}\)\right) \no \\
&=&{\cosh x\over N\pi}\left(
 \sum_{b=1}^{n-1}\sum_{q=1}^{l}L\({1\over A^{(b,q)}(-\infty)}\)
-\sum_{b=1}^{n-1}\sum_{q=1}^{l}L\({1\over A^{(b,q)}(\infty)}\)\right) \no \\
&=&{\cosh x\over N\pi}\left(
  \sum_{b=1}^{n-1}\left[\sum_{q=1}^{l-p-1}L\({1\over A^{(b,q)}(-\infty)}\)
  +L\({1\over A^{(b,p)}(-\infty)}\)\right.\right]               \no \\
&&\h \left.+\sum_{b=1}^{n-1}\sum_{q=p+1}^{l}L\({1\over A^{(b,q)}(-\infty)}\)
  -\sum_{b=1}^{n-1}\sum_{q=1}^{l}L\({1\over A^{(b,q)}(\infty)}\)\right) \;.
\ee
With the help of the
dilogarithm identity  the finite-size corrections can be expressed as
\be
\hs{-1}\ln\b^{(a,p)}(x)
&=&{\cosh x\over N\pi} \left(
  s(m',n,p)+s(m{'\hspace*{-1pt}'},n,h-n-p)-s(m,n,h-n)\;\right)\; \no \\
&=&{\pi\cosh x\over 6N}\({(n^2-1)(h-n-p)\over h-p}-{n(n^2-1)m{'\hspace*{-1pt}'}
     (m{'\hspace*{-1pt}'}+2)\over h-p} \no\\
&&+\;{(n^2-1)p\over n+p}-{n(n^2-1)m'(m'+2)\over n+p}\no\\
&&-\;{(n^2-1)(h-p)\over h}+{n(n^2-1)m(m+2)\over h} \no\\
&&+\;6(m'+m{'\hspace*{-1pt}'}-m)\left\lfloor{n\over 2}
  \right\rfloor\left\lfloor{n+1\over 2}\right\rfloor+6\Z \)\;.\label{corr}
\ee
For the excited states $m{'\hspace*{-1pt}'},m',m$ are greater than $1$
and also no longer independent. We suppose that
\be
m'=m-m{'\hspace*{-1pt}'}+n\left\lfloor{m-m{'\hspace*{-1pt}'}\over
p}\right\rfloor  \label{3m}
\ee
The relation (\ref{3m}) among $m{'\hspace*{-1pt}'},m',m$ has been shown for
$n=2$ in \cite{KlPe:92} and for $n\ge 2$ in
\cite{ChRa:89}. Using this relation and comparing  (\ref{corr}) with the
finite-size correction
\be
\ln\b^p(x)={\pi\over 6N}\(c-12(\Delta+\ol{\Delta})+(k+\ol{k})\)\cosh x
 +{\Got o\!}\({1\over N}\),
\ee
(\ref{corr}) yields the central charge
\be c={(n^2-1)p\over n+p}-{n(n^2-1)p\over h(h-p)}
\ee and the conformal weights
\be
\Delta_{t,s}={n(n^2-1)\{[ht-(h-p)s]^2-p^2\}\over 24ph(h-p)}- {n(n^2-1)\nu
(n\nu -
2p)\over 24p(p+n)}\;,
\ee where the exponents $s=m$ and $t=m{'\hspace*{-1pt}'}$ are integers
satisfying
\be 1\le s\le \lf{l\over n-1}\rf\;, \h 1\le t\le \lf{l-p\over n-1}\rf
 \;,\h p=1,2,\cdots,h-2
\ee and
\be
\nu:=(s-t)-\left\lfloor{s-t\over p}\right\rfloor p\;.
\ee The integers $k$ and $\ol{k}$ are such that
\be
k+\ol{k}&=& n(n^2-1)\left\lfloor{s-t\over
p}\right\rfloor\(n\left\lfloor{s-t\over p}\right\rfloor+2\)
+6n\left\lfloor{s-t\over p}\right\rfloor\left\lfloor{n\over
2}\right\rfloor\left\lfloor{n+1\over 2}\right\rfloor+6\Z \no\\
&=&0\h\mbox{mod $6$}
\ee
do not contribute to the central charge and conformal weights and hence can be
discarded.

\section{Discussion}
We have presented the results for the finite-size
corrections of transfer  matrices for fused $A^{(1)}_{n-1}$ models of
Jimbo, Miwa and Okado. These calculations generalize the evaluation of
the finite-size corrections of transfer matrices for the fused ABF
models given in \cite{KlPe:92} and the central charge calculation of
the $A^{(1)}_{n-1}$ models given in \cite{KNS:93b}.
In this paper we have studied the fused transfer  matrices with
rectangular Young diagrams
\smallskip
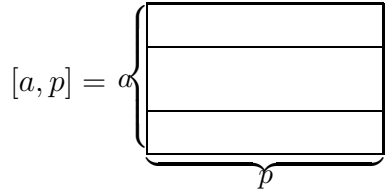
\begin{center}
$\cases{\hspace{-3pt}\setlength{\unitlength}{0.0075in}%
\begin{picture}(165,108.6)(75,660)
\put( 75,765){\line( 1, 0){165}}
\put(240,765){\line( 0,-1){105}}
\put(75,660){$\underbrace{\line(1, 0){165}}$}
\put( 75,660){\line( 0, 1){105}}
\put( 75,735){\line( 1, 0){165}}
\put( 75,690){\line( 1, 0){165}}
\put(55,707){\small$a$}
\put(153,640){\small$p$}
\put(-20,704){$[a,p]=$}
\end{picture}}$
\end{center}

\bigskip\bigskip
\noindent
These transfer matrices satisfy the functional
equations (\ref{func-T}). The finite-size corrections of the
transfer matrices with the fusion of Young diagram $[a,p]$ in
vertical direction are expressed by (\ref{corr}). Using Young
diagrams the result (\ref{corr}) can be represented by
\be
\;\sum_{b=1}^{n-1} \([b,l+n-p]+[b,p+n]-[b,l+n]\)
\label{A-Young}\ee
Suppose that the solution (\ref{su(n)2}) corresponds to
the Young diagram $[b,q]$. The expression (\ref{A-Young}) gives
the Rogers dilogarithm explanation for the central charges
(\ref{c-np}) and the  conformal weights (\ref{delta-np}),
which are missed in \cite{Kirillov:93}. The expression (\ref{A-Young})
also shows the coset structure corresponding to the central charge
(\ref{c-np}) and the conformal weights (\ref{delta-np1}) with the case
of (\ref{case}). For other cases the similar coset has been conjectured
\cite{ChRa:89} and need to be confirmed, which have not
been done in this paper.

The conformal weights obtained in this paper are a subset of all
possible
conformal weights (\ref{delta-np1}). But our calculation has shown
that in order to have  all conformal weights we have to analyze
the complex eigenvalues of the inversion identity hierarchies of the
models. The difficulty part is to find the related dilogarithm identity
involving the complex eigenvalues. These interesting questions are left
for further investigation.

The method by solving the functional equations of fusion hierarchies
to find the finite size corrections of the eigenvalues of the
fused transfer matrices firstly was developed in \cite{KlPe:92} for
study of the fused ABF restricted SOS models and more recently for
the central charges  of the fused $A_{n-1}^{(1)}$ models
\cite{KNS:93b}.  In fact, the functional
equations of the fused transfer matrices contain  enough information
to calculate  the eigenvalues of the transfer matrices. Hence the eigenvalues
and the relevant Bethe ansatz equations
can be extracted from the functional equations and
the finite size corrections can be calculated by solving
the Bethe ansatz equations \cite{BaRe:89,BaRe:90}. These two methods
technically are different, solving Bethe ansatz one employs the
string hypothesis and the other one relies on the ANZC property.
But it seems that by solving the functional equations it is more powerful
to find the conformal weights. Our present paper has shown this partially
for the fused $A_{n-1}^{(1)}$ models.
There are a large group of fused transfer matrices for the higher rank
$n>2$. For example, the fused transfer matrices with other Young diagrams.
These have not been evaluated here. But the finite-size corrections of these
transfer matrices should fit the picture (\ref{A-Young}).  However, we need
other functional equations to
solve the fused transfer matrices with other Young diagrams.

There are many solvable IRF models, e.g. $B_n^{(1)},C_n^{(1)},D_n^{(1)}$
models in \cite{JMO:88b}, $A_n^{(2)}$ models in \cite{Kuniba:91} and
dilute $A_L$ models in \cite{WNS:92}. Also we have the functional equations
for the fused transfer matrices of the models
\cite{KNS:93,KNS:93b,PeZh:94,ZPG:94,Zhou:94}. It should be possible
to solve these functional equations by similar methods.

\bigskip

 \section*{Acknowledgements}  This research is supported by the Australian
Research Council.  The authors thank B. Y. Hou and Ole Warnaar
for discussions.

\bigskip\newpage

\appendix
\subsection*{Appendix~A: The adjacency matrices of the fused JMO
$A^{(1)}_{n-1}$ models for
$n=4$ and $l=2$.}
\setcounter{equation}{0}
\renewcommand{\theequation}{A.\arabic{equation}} The set of dominant integrable
weights $P_+(4,2)$ is given by [(0,0,0),(1,0,0),(0,1,0),
(0,0,1),(2,0,0),(1,1,0),(1,0,1),(0,2,0),(0,1,1),(0,0,2)] according to the
Young diagram notation (see Figure~\ref{fig2}). The adjacency matrices of
the model  with rank
$n=4$ and level
$l=2$ is labeled by the order of the elements in $P_+(4,2)$ for the rows and
the
columns of matrix.

\be A^{(0,0,0)}&=& 1\no \\ A^{(m,n,l)}&=&0 \h \mbox{for $m+n+l>2$ and
$m,n,l<0$} \no
\ee

\bigskip
\twocol
\hfil{$A^{(0,0,1)}=$}\hfil&\hfil{$A^{(0,1,0)}=$}\hfil\\
\nopagebreak\smallskip\nopagebreak
\twocol
$\smat{0&0&0&1&0&0&0&0&0&0\cr
 1&0&0&0&0&0&1&0&0&0\cr
 0&1&0&0&0&0&0&0&1&0\cr
 0&0&1&0&0&0&0&0&0&1\cr
 0&1&0&0&0&0&0&0&0&0\cr
 0&0&1&0&1&0&0&0&0&0\cr
 0&0&0&1&0&1&0&0&0&0\cr
 0&0&0&0&0&1&0&0&0&0\cr
 0&0&0&0&0&0&1&1&0&0\cr
 0&0&0&0&0&0&0&0&1&0}$&
$\smat{0&0&1&0&0&0&0&0&0&0\cr
 0&0&0&1&0&1&0&0&0&0\cr
 1&0&0&0&0&0&1&1&0&0\cr
 0&1&0&0&0&0&0&0&1&0\cr
 0&0&0&0&0&0&1&0&0&0\cr
 0&1&0&0&0&0&0&0&1&0\cr
 0&0&1&0&1&0&0&0&0&1\cr
 0&0&1&0&0&0&0&0&0&0\cr
 0&0&0&1&0&1&0&0&0&0\cr
 0&0&0&0&0&0&1&0&0&0}$\\
\twocol
\hfil{$A^{(0,1,1)}=$}\hfil&\hfil{$A^{(1,0,0)}=$}\hfil\\
\nopagebreak\smallskip\nopagebreak
\twocol
$\smat{0&0&0&0&0&0&0&0&1&0\cr
 0&0&1&0&0&0&0&0&0&1\cr
 0&0&0&1&0&1&0&0&0&0\cr
 0&0&0&0&0&0&1&1&0&0\cr
 0&0&0&1&0&0&0&0&0&0\cr
 1&0&0&0&0&0&1&0&0&0\cr
 0&1&0&0&0&0&0&0&1&0\cr
 0&1&0&0&0&0&0&0&0&0\cr
 0&0&1&0&1&0&0&0&0&0\cr
 0&0&0&0&0&1&0&0&0&0}$&
$\smat{0&1&0&0&0&0&0&0&0&0\cr
 0&0&1&0&1&0&0&0&0&0\cr
 0&0&0&1&0&1&0&0&0&0\cr
 1&0&0&0&0&0&1&0&0&0\cr
 0&0&0&0&0&1&0&0&0&0\cr
 0&0&0&0&0&0&1&1&0&0\cr
 0&1&0&0&0&0&0&0&1&0\cr
 0&0&0&0&0&0&0&0&1&0\cr
 0&0&1&0&0&0&0&0&0&1\cr
 0&0&0&1&0&0&0&0&0&0}$\\
\bigskip
\twocol
\hfil{$A^{(1,1,0)}=$}\hfil&\hfil{$A^{(1,0,1)}=$}\hfil\\
\nopagebreak\smallskip\nopagebreak
\twocol
$\smat{0&0&0&0&0&1&0&0&0&0\cr
 0&0&0&0&0&0&1&1&0&0\cr
 0&1&0&0&0&0&0&0&1&0\cr
 0&0&1&0&1&0&0&0&0&0\cr
 0&0&0&0&0&0&0&0&1&0\cr
 0&0&1&0&0&0&0&0&0&1\cr
 0&0&0&1&0&1&0&0&0&0\cr
 0&0&0&1&0&0&0&0&0&0\cr
 1&0&0&0&0&0&1&0&0&0\cr
 0&1&0&0&0&0&0&0&0&0}$&
$\smat{0&0&0&0&0&0&1&0&0&0\cr
 0&1&0&0&0&0&0&0&1&0\cr
 0&0&1&0&1&0&0&0&0&1\cr
 0&0&0&1&0&1&0&0&0&0\cr
 0&0&1&0&0&0&0&0&0&0\cr
 0&0&0&1&0&1&0&0&0&0\cr
 1&0&0&0&0&0&1&1&0&0\cr
 0&0&0&0&0&0&1&0&0&0\cr
 0&1&0&0&0&0&0&0&1&0\cr
 0&0&1&0&0&0&0&0&0&0}$\\
\bigskip
\goodbreak
\twocol
\hfil{$A^{(2,0,0)}=$}\hfil&\hfil{$A^{(0,2,0)}=$}\hfil\nopagebreak\\
\nopagebreak\smallskip\nopagebreak
\twocol
$\smat{0&0&0&0&1&0&0&0&0&0\cr
 0&0&0&0&0&1&0&0&0&0\cr
 0&0&0&0&0&0&1&0&0&0\cr
 0&1&0&0&0&0&0&0&0&0\cr
 0&0&0&0&0&0&0&1&0&0\cr
 0&0&0&0&0&0&0&0&1&0\cr
 0&0&1&0&0&0&0&0&0&0\cr
 0&0&0&0&0&0&0&0&0&1\cr
 0&0&0&1&0&0&0&0&0&0\cr
 1&0&0&0&0&0&0&0&0&0}$&
$\smat{0&0&0&0&0&0&0&1&0&0\cr
 0&0&0&0&0&0&0&0&1&0\cr
 0&0&1&0&0&0&0&0&0&0\cr
 0&0&0&0&0&1&0&0&0&0\cr
 0&0&0&0&0&0&0&0&0&1\cr
 0&0&0&1&0&0&0&0&0&0\cr
 0&0&0&0&0&0&1&0&0&0\cr
 1&0&0&0&0&0&0&0&0&0\cr
 0&1&0&0&0&0&0&0&0&0\cr
 0&0&0&0&1&0&0&0&0&0}$\\
\bigskip
\twocol
\hfil{$A^{(0,0,2)}=$}\hfil&\hfil{}\hfil\\
\nopagebreak\smallskip\nopagebreak
\twocol
$\smat{0&0&0&0&0&0&0&0&0&1\cr
 0&0&0&1&0&0&0&0&0&0\cr
 0&0&0&0&0&0&1&0&0&0\cr
 0&0&0&0&0&0&0&0&1&0\cr
 1&0&0&0&0&0&0&0&0&0\cr
 0&1&0&0&0&0&0&0&0&0\cr
 0&0&1&0&0&0&0&0&0&0\cr
 0&0&0&0&1&0&0&0&0&0\cr
 0&0&0&0&0&1&0&0&0&0\cr
 0&0&0&0&0&0&0&1&0&0}$&
${}$\\

\bigskip



\begin{thebibliography}{99}
\bibitem{BPZ:84} A. A. Belavin, A. M. Polyakov and A. B. Zamolodchikov,
    Nucl Phys. \newline {\bf B 241}(1984)333.
\bibitem{FQS:84}D. Friedan, Z. Qiu and S. Shenker, Phys. Rev. Lett.
   {\bf 52}(1984)1575.
\bibitem{ABF:84} G. E. Andrews, R. J. Baxter and P. J. Forrester, J. Stat.
   Phys. {\bf 35} (1984) 193.
\bibitem{Huse:84}D. A. Huse, Phys. Rev. {\bf B30}(1984)3908.
\bibitem{Cardy:86} J. L. Cardy, Nucl Phys.{\bf B270}(1986) 186.
\bibitem{Kac:79} V. G. Kac, Lecture notes in physics {\bf 94}(1979) 441.
\bibitem{BaRe:89} V. V. Bazhanov and N. Yu Reshetikhin, Int. J. Mod. Phys.
   {\bf B4} (1989) 115.
\bibitem{KlPe:92} A. Kl{\"{u}}mper and P. A. Pearce, Physica A {\bf 183}
     (1992) 304.
\bibitem{DJMO:87} E. Date, M. Jimbo, T. Miwa and M. Okado, Phys. Rev. {\bf
   B35} (1987) 2105.
\bibitem{DJKMO:87} E. Date, M. Jimbo, A. Kuniba, T. Miwa and M. Okado,
           Nucl Phys. {\bf B290} (1987) 231.
\bibitem{DJKMO:88} E. Date, M. Jimbo, A. Kuniba, T. Miwa and M. Okado,
           Adv. Stud. Pure Math.,{\bf 16} (1988) 17.
\bibitem{GKO:85} P.Goddard, A.Kent, D.Olive, Phys. Lett., {\bf B152},
(1985) 105;
                Commun. Math. Phys. {\bf 103} (1986) 105.
\bibitem{FaLu:88} V. Fateev and S. L. Lukyanov, Int. J. Mod. Phys.
   {\bf A3}(1988) 507.
\bibitem{BBSS:88}B. Bais, P. Bouwknegt, K. Schoutens and M. Surridge,
                  Nucl. Phys. {\bf B304}(1988) 348; 371.
\bibitem{JMO:88} M. Jimbo, T. Miwa and M. Okado, Nucl. Phys. {\bf
   B300} (1988) 74.
\bibitem{ChRa:89} P.Christe and F.Ravanini, Int. J. Mod. Phys., {\bf A4},
(1989) 897.
\bibitem{BaRe:90} V. V. Bazhanov and N. Yu Reshetikhin, J. Phys.
     {\bf A23} (1990) 1477.
\bibitem{KNS:93}A. Kuniba, T. Nakanishi and J. Suzuki, ``Functional relations
     in solvable lattice models I: functional relations and representation
       theory", to be published in Int. J. Mod. Phys. {\bf A}.
\bibitem{KNS:93b}A. Kuniba, T. Nakanishi and J. Suzuki, ``Functional relations
    in solvable lattice models II: applications", to be published in
       Int. J. Mod. Phys. {\bf A}.
\bibitem{KuNa} A.Kuniba and T.Nakanishi,
  Rogers dilogarithm in integrable systems, preprint SMS-92/A046, 1992.
\bibitem{FeFu:82} B. L. Feigin and K. B. Fuchs, Funct. Anal. Appl. {\bf
16}(1982) 114.
\bibitem{DoFa:84}Vl. S. Dotsenko and V. A. Fateev, Nucl. Phys. {\bf
240}(1984) 312.
\bibitem{Kirillov:93} A.N.Kirillov, Dilogarithm identities and spectra in
conformal
   field theory, Talk given at the Isaac Newton Institute, Cambridge,
   October 1992, preprint hep-th/9211137.
\bibitem{DeWo:85}H. J. de Vega and F. Woynarovich, Nucl Phys. {\bf
B251}(1985) 439.
\bibitem{Wo:87}F. Woynarovich, Phys. Rev. Lett. {\bf 59}(1987) 259.
\bibitem{GeRi:87}G. von Genlen and V. Rittenberg, J. Phys. {\bf A20}(1987) 227.
\bibitem{DeKa:87}H. J. de Vega and M. Karowski,  Nucl Phys. {\bf
B285}(1987) 619.
\bibitem{Ka:88}M. Karowski,  Nucl Phys. {\bf B300}(1988) 473.
\bibitem{KiRe:87} A.N.Kirillov, N.Yu.Reshetikhin, J. Phys. A. Math. Gen.,
          {\bf 20} (1987) 1565.
\bibitem{ABB:87}F. C. Alcaraz, M. N. Barber and M. T. Batchelor, Phys. Rev.
          Lett.{\bf 58}(1987) 771.
\bibitem{BNW:89}M. T. Batchelor, B. Niehuis and S. O. Warnaar, Phys. Rev.
                 Lett. {\bf 62}(1989) 2425.
\bibitem{KlPe:91}A. Kl{\"u}mper and P. A. Pearce, J. Stat. Phys. {\bf
64}(1991) 13.
\bibitem{BCN:86}H. W. J. Bl{\"{o}}te, J. L. Cardy and M. P. Nightingale, Phys.
                Rev. Lett. {\bf 56}(1986) 742.
\bibitem{Affleck:86}I. Affleck Phys. Rev. Lett. {\bf 56}(1986) 746.
\bibitem{JMO:87} M. Jimbo, T. Miwa and M. Okado, Lett. Math. Phys. {\bf
   14} (1987) 123.
\bibitem{JKMO:88} M. Jimbo, A. Kuniba, T. Miwa and M. Okado, Commun. Math.
Phys.
    {\bf 119} (1988) 543.
\bibitem{ZhPe:94} Y. K. Zhou and P. Pearce, Fusion and \ade lattice models,
      Int. J. Mod. Phys (1994).
\bibitem{SaAt:91} H. Saleur and D. Altshuler, Nucl. Phys. {\bf B354}(1991) 579.
\bibitem{KuNa:89} A.Kuniba, T.Nakanishi, Level-rank duality in fusion RSOS
model, prepr. 1989.
\bibitem{JMO:88b} M. Jimbo, T. Miwa and M. Okado, Commun. Math. Phys. {\bf
   116} (1988) 507.
\bibitem{Kuniba:91}A. Kuniba, Nucl. Phys. {\bf B355}(1991) 801.
\bibitem{WNS:92} S. O. Warnaar, B. Nienhuis and K. A. Seaton, Phys. Rev.
Lett. {\bf 69} (1992) 710.
\bibitem{PeZh:94} P. A. Pearce and Y-K Zhou, ``Yang-Baxter Algebras and Fusion
   of $A$--$D$--$E$ Lattice Models", to be published in Int. J. Mod. Phys.
   {\bf B} (1994).
\bibitem{ZPG:94}Y. K. Zhou,  P. A. Pearce and U. Grimm,
                ``Fusion of Dilute A Lattice Models", preprint(1994).
\bibitem{Zhou:94}Y. K. Zhou, ``$SU$(2) hierarchies of dilute lattice
                models", preprint(1994).




\end{thebibliography}
\end{document}